\documentstyle[12pt,aaspp]{article}

\lefthead{PAGE}
\righthead{MAGNETIZED NEUTRON STAR TEMPERATURE}

\newcommand{\tnm}{\tablenotemark}
\newcommand{\tnt}{\tablenotetext}
\newcommand{\be}{\begin{equation}}
\newcommand{\ee}{\end{equation}}
\newcommand{\etal}{{\em et al. }}
\newcommand{\ba} {$\begin{rm}\begin{array}{c} }
\newcommand{\bba} {$\left\{ \begin{rm}\begin{array}{c} }
\newcommand{\baa}{$\begin{rm}\begin{array}{cc}}
\newcommand{\ea} {\end{array}\end{rm}$}
\newcommand{\eab} {\end{array}\end{rm} \right.$}
\newcommand{\nal}{\protect\\}
\newcommand{\LM}{{\rm LM}}
\newcommand{\R}{{\rm R}}
\newcommand{\M}{{\rm M}}

\def\gsim{\lower 2pt \hbox{$\, \buildrel {\scriptstyle >}\over
         {\scriptstyle \sim}\,$}}
\def\lsim{\lower 2pt \hbox{$\, \buildrel {\scriptstyle <}\over
         {\scriptstyle \sim}\,$}}

\newcommand{\Msol}{\mbox{$M_{\odot}\;$}}

\begin{document}

\title{SURFACE TEMPERATURE OF A MAGNETIZED NEUTRON STAR AND
       INTERPRETATION OF THE {\em ROSAT} DATA. I. DIPOLAR FIELDS}
\author{Dany Page}
\affil{Instituto de Astronom\'{\i}a, UNAM, Apdo Postal 70-264,
       04510 M\'{e}xico D.F., M\'{e}xico
       \footnote{Current address.}\\
       Department of Astronomy, Columbia University, New York, NY 10027, USA}

\begin{abstract}

We model the temperature distribution at the surface of a magnetized
neutron star and study the effects on the observed X-ray spectra and light
curves.
General relativistic effects, i.e., red-shift and lensing, are fully taken
into account.
Atmospheric effects on the emitted spectral flux are
not included: we only consider blackbody emission at the local
effective temperature.
In this first paper we restrict ourselves to dipolar fields.
General features are studied and compared with the {\em ROSAT} data
from the pulsars 0833-45 (Vela), 0656+14, 0630+178 (Geminga), and
1055-52, the four cases for which there is strong evidence that
thermal radiation from the stellar surface is detected.

The composite spectra we obtain are not very different from a
blackbody spectrum at the star's effective temperature.
We conclude that, as far as blackbody spectra are considered,
temperature estimates using single temperature models give results
practically identical to our composite models.
The change of the (composite blackbody) spectrum with the star's
rotational phase is also not very large and may be unobservable in
most cases.

Gravitational lensing strongly suppresses the light curve pulsations.
If a dipolar field is assumed, pulsed fractions comparable to the
observed ones can only be obtained with stellar radii larger than what
predicted by current models of neutron star structure, or with low
stellar masses.
Moreover, the shapes of the theoretical light curves with dipolar
fields do not correspond to the observations.
The use of magnetic spectra may rise the pulsed fraction sufficiently,
but will certainly make the discrepancy with the light curve shapes
worse:
dipolar field are not sufficient to interpret the data.
Many neutron star models with a meson condensate or hyperons predict
very small radii, and hence very strong lensing, which will require
highly non dipolar fields to be able to reproduce the observed pulsed
fractions, if possible at all:
this may be a new tool to constrain the size of neutron stars.

The pulsed fractions obtained in all our models increase with photon
energy:
the strong decrease observed in Geminga at energies 0.3 - 0.5 keV is
definitely a genuine effect of the magnetic field on the spectrum in
contradistinction to the magnetic effects on the surface temperature
considered here.
Thus, a detailed analysis of thermal emission from the four pulsars we
consider will require both complex surface field configurations and
the inclusion of magnetic effects in the atmosphere (i.e., on the
emitted spectrum).

\end{abstract}

\keywords{\quad dense matter \quad ---
          \quad stars: neutron \quad ---
          \quad stars: X-rays}

\bigskip
\bigskip
Submitted to {\em The Astrophysical Journal}

\newpage
\section{INTRODUCTION}

The detection of thermal emission from the surface of a neutron star is
one of the `Holy Graal's of X-ray astronomy.
At the first detection of X-rays from the direction of the Crab
supernova remnant, surface thermal radiation from the neutron star
already expected to be present in this remnant was proposed as the
most probable source (Bowyer \etal 1964), but the claim was soon
disproved and even the most recent {\em ROSAT} observation of the Crab
pulsar failed to detect any emission from the surface
(Becker \& Aschenbach 1993).
A compilation of all {\em Einstein} observations (Seward \& Wang 1988)
listed ten neutron stars detected in soft X-rays, to which should now
be added Geminga which at that time was not yet proved to be a neutron
star but had been clearly seen by {\em Einstein} (Bignami, Caraveo \&
Lamb
1983).
As of August 1993 there were thirteen confirmed detections of pulsars
by {\em ROSAT} and six unconfirmed ones (\"{O}gelman 1993).
In most cases the radiation is probably of magnetospheric origin with some
contamination from the surrounding synchrotron nebula, at least for the
younger candidates, and surface thermal emission in some cases.
With the sensitivity of the {\em ROSAT}'s PSPC (Position Sensitive
Proportional Counter) and long exposure times there is now strong
spectral evidence that thermal radiation has been detected from four
neutron stars (\"{O}gelman 1993):
PSR 0833-45 (Vela) (\"{O}gelman, Finley  \& Zimmermann 1993),
PSR 0656+14 (Finley, \"{O}gelman \& Kizilo\u{g}lu 1992),
PSR 0630+178 (Geminga) (Halpern \& Holt 1992), and
PSR 1055-52 (\"{O}gelman \& Finley 1993).
Moreover, these four objects show pulsed X-ray emission, three of them
(except Geminga) are radio pulsars and three (except PSR 0656+14) have
also been detected as $\gamma$-ray pulsars.

Earlier {\em Einstein} observations had already provided some limited
evidence for detection of surface thermal radiation in the cases of
the Vela pulsar (Harnden \etal 1985), PSR 0656+14 (C\'{o}rdova \etal
1989), and Geminga (Halpern \& Tytler 1988), while in the case of PSR
1055-52 (Cheng \& Helfand 1983) thermal emission was considered as
incompatible with the data.
In later {\em EXOSAT} observations, with broad band spectroscopy only,
thermal radiation was considered as the most reasonable origin of the
detected soft X-rays in both cases of the Vela pulsar (\"{O}gelman
\& Zimmermann 1989) {\em and} PSR 1055-52 (Brinkmann \&
\"{O}gelman 1987).
Because of its late discovery, as an X-ray source, in the {\em Einstein}
data base, PSR 0656+14 has not been observed by {\em EXOSAT}.
The {\em EXOSAT} observation of Geminga (Caraveo \etal 1984) did not
give any new spectral information compared to the {\em Einstein} results.
A review of the pre-{\em ROSAT} observational situation has been given by
\"{O}gelman (1991).

The quality of the {\em ROSAT} data from these four nearby neutron stars
presents a new chalenge for theorists to provide good models for their
interpretation.
The heretofore published analyses of these data have all assumed a
unique surface temperature, with at most a second thermal component
coming from the small hot polar caps.
We model here the temperature distribution at the surface of a magnetized
neutron star and study its effects on the received spectra and
light curves.
The crustal magnetic field affects the heat transport in the layers
beneath the surface and makes that regions of the star where the field is
almost normal to the surface will be warmer than regions where the
field is almost tangential to the surface.
These temperature differences will give rise to modulation of the
received flux at energies between 0.1 and 1 keV and are a natural
explanation for the observed pulsations in the above mentioned four
neutron stars.
This long searched for detection of thermal radiation from the surface
of neutron stars thus opens up a new window in the study of these objects.
It has the potential to tell us about the structure of the surface
magnetic field and give us new information about the size of these stars
through gravitational lensing effects which obviously will be substantial.
Our purpose in this first paper is to present the general physics
involved and study the simplest case of a dipolar surface magnetic
field.
We have tried to present as clearly as possible the underlying
physical ingredients as well as the method used, laying hopefully
a clear groundstone for future improvements and/or modifications.

General relativistic effects may be enormous in neutron stars and
we take fully into account both gravitational red-shift and
gravitational lensing.
Magnetic fields are also affected by strong gravitational fields but
we do not consider this effect:
of critical importance for the surface temperature distribution is the
angle between the surface's normal and the magnetic field, and this
angle is practically unaffected by gravity (Ginzburg \& Ozernoy 1964).
The major effect of gravity on the surface magnetic field is to
increase its strength at small radii:
for a given dipolar field at infinity the surface strength is
increased by 20-50\% by gravity compared to the value it would have in
flat space-time.
The surface field strengths we consider should thus simply be somewhat
reduced for comparison with values obtained for example from pulsar
spin-down.
Our results are however not very sensitive to changes of this size in
the field strength.
We do not include the effects of the magnetic field on the
atmosphere where the emitted spectrum is determined.
These effects are also substantial and will hopefully be included
in future work.

It has been proposed that the surface of neutron stars may be a
magnetic solid (Ruderman 1974; Chen, Ruderman \& Sutherland 1974).
Even if improved calculations of the atomic lattice cohesive energy in
very strong magnetic fields have indicated that this is not the case
(Jones 1986), one should nevertheless keep in mind this possibility.
It would obviously have dramatic effects on the emitted spectrum
(Brinkmann 1980) and some, comparatively smaller, effect on the
surface temperature (see Van Riper 1988 for simple estimates)
Our model is based on magnetized envelope calculations which assume that
the surface
is not a magnetic solid:
would this be uncorrect, our results should clearly be reconsidered.

We must also mention that the interpretation presented here assumes
that the surface temperature is determined by the flow of heat from the
star's interior and thus carries the imprint of the underlying crustal
magnetic field.
Another interpretation has been proposed (Halpern \& Ruderman 1993) in
which the hard X-rays emitted by the hot polar caps are scattered back
onto the surface by the magnetospheric plasma:
if this is the case the surface temperature has nothing to do with the
properties of the magnetized envelope and our model is then
obviously irrelevant.
A composite model is also possible: the general temperature
distribution may be determined by the heat flow from the interior and
some regions heated by the back scattered hard X-rays.
Moreover, absorption and scattering of radiation by the surrounding
magnetospheric plasma may also be an important factor in reshaping the
emitted flux (Halpern \& Ruderman 1993).
Which of these possibilities is actual can only be determined by
studying each of them carefully.
This paper is a first step in that direction.

The structure of the paper is a follow.
The {\em ROSAT} data are summarized in \S~\ref{sec:data} where we
emphasize the features relevent to our purpose.
In \S~\ref{sec:tbts} we describe the effects of the magnetic field on
heat transport in the envelope and our model for the surface
temperature distribution and in \S~\ref{sec:litc_sp} we present the
method used to calculate the fluxes as observed by {\em ROSAT}.
Our results are in \S~\ref{sec:results} and they are discussed in
\S~\ref{sec:disc}, followed by our conclusions in \S~\ref{sec:concl}

\section{THE {\em ROSAT} DATA \label{sec:data}}

We present in this section some characteristics of the {\em ROSAT}
data for the four neutron stars we study along with some relevant
informations obtained at other energies, summarized in
Table~\ref{tab:data}.
Vela is the younger of the four and an accurate estimate of its
surface temperature is of outmost importance since it could give us
evidence for the occurence of fast neutrino emission in its early
cooling history (Page \& Applegate 1992).
The other three are much older and their importance for cooling is
distinct; they do provide us with insight about the state of their
interior (Page 1994a).
All four are X-ray pulsars, three of them (except Geminga) are also
radio pulsars, three (except PSR 0656+14; see however Ma \etal 1993 and
Brown \& Hartmann 1993) are $\gamma$-ray pulsars and
three (except PSR 1055-52) have an optical counterpart.
Analysis of the X-ray emission from the Vela pulsar is delicate
because of the strong emission from the surrounding synchrotron nebula
which has to be substracted from the data to obtain the signal from
the pulsar itself.
The other three X-ray sources are much cleaner than Vela:
there is no evidence of extended emission in the {\em ROSAT} HRI
observation of PSR 0656+14 (Anderson \etal 1993) and PSR 1055-52
(\"{O}gelman \& Finley 1993).
There is no HRI observation of Geminga to date but the PSPC data are
compatible with a point source.

All four objects show pulsed X-ray emission within the {\em ROSAT}
energy range (0.08 - 2.5 keV) and a two component spectrum, the soft
component (roughly below 0.5 keV) being interpreted as surface thermal
emission (\"{O}gelman 1993).
The corresponding pulsed fractions of these soft components are listed in
Table~\ref{tab:data}.
Geminga's pulsed fraction below 0.28 keV is larger than between 0.28 - 0.5 keV
while the other three pulsars have a roughly constant pulsed fraction below
0.5 keV (a slight decrease seems to be present in PSR 1055-52 below 0.5 keV).
The phase of the peak in the light curve of these soft components is
independent of the photon energy; the peaks are broad, the flux being
above its mean value more than 50\% of the time; the light curves are
non sinusoidal.
The hard component of Vela appears to be unpulsed: it may be real or just
due to the fact that count rates are low in this part of the spectrum and
the signal has been extracted from the nebula emission.
In the case of PSR 0656+14 the count rate is too low to detect pulsation
of the hard component but in the other two cases (Geminga and PSR 1055-52)
pulsations are clearly seen and they are substantially off-phase compared
to the soft component.
The origin of the hard component of these spectra still remains ellusive
due to the low count rates, fits with blackbody or power-law spectra
giving equally good (or bad) results.
It could be thermal emission from the hot polar caps or some magnetospheric
process coming maybe, but not necessarily, from just above the polar caps.
In this paper `soft' and `hard' X-rays will always refer to these two
components within the {\em ROSAT} energy range.

The effective temperatures $T_e^{BB}$ and interstellar column
densities $N_H^{BB}$ listed in Table~\ref{tab:data} were obtained
from single temperature blackbody fits.
In the case of Vela, the quoted effective temperature and the observed
flux imply a stellar radius of 3-4 km, i.e., only 10\% of the surface
is emitting at this temperature:
this may be due to very large surface temperature gradients or to a
strong inadequacy of the blackbody spectrum.
Use of more complex spectra from non magnetized He atmosphere (Romani
1987) or magnetized H atmosphere (Pavlov \etal 1994) give lower
temperatures: $2.2 - 2.6 \times 10^5$ K (Finley \etal 1992) and $6.6 -
7.4 \times 10^5$ K (Anderson \etal 1993) respectively instead of $8.6
- 9.4 \times 10^5$ K with a blackbody spectrum in the case of PSR
0656+14.
In the case of Geminga, magnetized hydrogen atmosphere spectra also
give lower temperatures, $2 - 3 \times 10^5$ K (Meyer \etal 1994).

The (dipolar) magnetic field strengths we cite are obtained the
standard way (magnetic dipole radiation braking) by
\be
B_p
    = 2 \times 10^{12} (P \dot{P}_{-15})^{1/2} {\rm G}
\label{equ:B_p}
\ee
where $P$ is the period in seconds and $\dot{P}_{-15}$ its derivative in
units of
$10^{-15} \; {\rm s \; s^{-1}}$.
$B_p$ is thus the field strength at the magnetic pole on the  star's
surface.
These values are of course only indicative, and general relativistic
strengthening of the surface field adds to this uncertainty.
Phenomenological analyses of the radio data allow to estimate the
inclination angle $\alpha$ of the dipole with respect to the rotation
axis and the angle $\zeta$ between the observer's direction and the
rotation axis, but the results are strongly model dependent in many
cases (Miller \& Hamilton 1993).
PSR 1055-52 is generally considered as a typical example of radio
inter-pulsar where emission from both magnetic poles is detected
($\alpha \sim \zeta \sim 90^0$);
an alternative explanation for the interpulse is that the same
magnetic pole is seen twice per rotation ($\alpha \sim \zeta \sim
0^0$).
In the case of Geminga, modeling of the $\gamma$-ray emission within
both the outer gap (Halpern \& Ruderman 1993) and polar cap (Harding,
Ozernoy \& Usov 1993) models indicate that $\alpha \sim \zeta \sim
90^0$.
The beaming of radio emission obviously requires $\alpha \sim \zeta$
for radio pulsars.

\section{STRUCTURE OF THE ENVELOPE \label{sec:tbts}}

For all neutron stars we are presently interested in, the interior is
isothermal and a temperature gradient only appears in the upper layers
of the outer crust (Nomoto \& Tsuruta 1986, 1987).
We will call these layers the {\em envelope} and fix its upper density
at $\rho = 10^{10} \; {\rm gm/cm^3}$.
On top of the envelope resides the {\em atmosphere}: the total energy
flux reaching it is determined within the underlying envelope, mostly
by electron transport, and cannot be changed anymore, but its energy
distribution, i.e., the emitted spectrum, is determined here.

\subsection{Envelope without magnetic field \label{sec:tbts_noH}} 

Detailed calculations of heat transport in non magnetized neutron star
envelopes have been presented by Gudmundsson, Pethick \& Epstein
(1982, 1983) and Hernquist \& Applegate (1984)
These authors noticed that the study of the envelope can be separated
from the study of the interior, since for neutron stars several months
old the evolution time scale of the interior is much larger that the
thermal time scale of the envelope, and the latter can be considered to be
in a steady state.
The structure of the envelope depends on the gravitational
acceleration at the surface
\be
g_s = \frac{GM}{R^2 \sqrt{1-2GM/Rc^2}}
\ee
and on the temperature at its base $T_b$, i.e., at the
envelope-interior boundary at density $\rho = 10^{10} {\rm gm \;
cm^{-3}}$.
Since the interior is isothermal its temperature is $T_{int} = T_b$.
The envelope then determines the effective temperature $T_e$ related
to the interior temperature $T_b$ quite accurately by
\be
T_{b 8} \cong 1.3 \; \left(\frac{T_{e 6}^4}{g_{s 14}}\right)^{0.455}
\ee
(Gudmundsson, Pethick \& Epstein 1982) where $X_n$ is $X$ measured in
$10^n$ cgs units.
Since the interior is isothermal and, in absence of magnetic field,
the heat transport in the envelope is isotropic the resulting surface
temperature is uniform over the star: $T_s(\theta,\phi) = T_e$.

The heat transport is due to electrons in the lower layers and to
photons in the upper layers.
The critical region, ``sensitivity strip'', which determines the $T_b
- T_e$ relationship is the region just below the electron-photon
transport transition where heat is thus transported mainly be
electrons and the ions are in the liquid phase (Gudmundsson, Pethick
\& Epstein 1982, 1983; Hernquist \& Applegate 1984).
When the surface temperature decreases below about $\sim 3 \cdot 10^5$
K, the sensitivity strip is mostly in a region where the ions are
partially ionized and there is no reliable calculation of the electron
conductivity in this regime.
Electron scattering by impurities also begins to contribute
significantly; it depends on the impurity concentration, at most
a poorly known quantity.
Moreover the photon opacity has to be extrapolated significantly from
calculated values.
This makes that the surface temperatures cannot be reliably calculated
and all such values ($T_s < 3 \cdot 10^5$ K) are only illustrative.
The same remark apply {\em a fortiori} to the magnetic cases
considered in the next subsection.

\subsection{Envelope with magnetic field \label{sec:tbts_H}}

In presence of a strong magnetic field the electron motion in
directions perpendicular to the field becomes quantized, the electrons
occupying Landau levels with spacings (non relativistic case)
\be
\hbar \omega_B = \frac{\hbar |e| B}{m_e c} =
                 11.3 \; {\rm keV} \; B_{12}
\ee
For typical fields present in neutron star envelopes electrons will
occupy, at zero temperature, only the first Landau level up to
densities of the order of $5 \cdot 10^5 {\rm gm \; cm^{-3}}$ and then
start to populate higher levels.
Temperature can push electrons into higher levels only when
\be
k_B T \gg \hbar \omega_B.
\ee
The structure of the electron Fermi surface is thus strongly affected
by the magnetic field almost across the whole sensitivity strip, and
the transport properties will consequently depend on the field.
Electron transport in directions perpendicular to the field is
strongly suppressed while it is slightly enhanced parallelly to the
field.
Photon opacities are also affected but the Rosseland mean, the
relevant quantity for heat transport, is only slightly anisotropic:
anisotropy in heat transport thus comes mainly from electrons.
A review of magnetized envelopes can be found in Yakovlev \& Kaminker
(1994).

When compared to the non magnetic case, the surface temperature will
be slightly higher if the magnetic field is radial (parallel
transport) but much lower if the magnetic field is tangential to the
surface (orthogonal transport).
The envelope being at most a few tens of meters thick one can safely
assume that the magnetic field is uniform across it.
Detailed modelling of magnetized neutron star envelopes with uniform
field have been presented by Hernquist (1985), Van Riper (1988) and
Schaaf (1990a) for the case of parallel transport while the only
published models for orthogonal transport are given by Schaaf (1990a).
Hernquist's (1985) and Van Riper's (1988) results are in good
agreement except at low temperature where Van Riper's inclusion of the
(negative) ion Coulomb pressure (omitted by Hernquist) showed that the
magnetic effects on the equation of state become dominant (Schaaf's
results are not in such good agreement due probably to the many
approximations done).
For the relevant field strengths, these magnetic $T_b - T_s$
calculations can hence only be trusted at $T_s \gsim 3 \cdot 10^5$ K,
the same range of validity as for the non magnetic case.
We will use Hernquist's (1985) results for parallel transport and
Schaaf's (1990a) results for orthogonal transport.
Figure \ref{fig:tbts} shows the resulting $T_b - T_s$ relationships.

In the particular cases of transport with a radial or a tangential
field the problem is one dimensional since the heat flux can only be
radial by symmetry.
For an arbitrary orientation of the field with respect to the radial
direction the problem of heat transport becomes two-dimensional.
However, if the field orientation is sufficiently uniform on a length
scale of a few tens of meters (= thickness of the envelope) the
problem can locally be reduced to one dimension, at least in a first
approximation.
The surface temperature $T_s$ then depends only on the field strength
$B$ and the angle $\Theta_B$ between the field and the radial
direction (beside the dependence on the envelope's bottom temperature
$T_b$ and the surface gravitational acceleration $g_s$).
Greenstein \& Hartke (1983) showed that one can then approximately
write (keeping only the $\Theta_B$ dependence explicit)
\be
T_s(\Theta_B) = \chi(\Theta_B) \times T_s(\Theta_B=0)
\label{equ:TsB}
\ee
where
\be
\chi(\Theta_B) =
   \left[\cos^2 \Theta_B + \chi_0^4 \sin^2 \Theta_B \right]^{1/4}.
\label{equ:chi}
\ee
In Greenstein \& Hartke's argument $\chi_0^4
=K_{\perp}/K_{\parallel}$, the ratio of the thermal conductivities
perpendicular and parallel to the magnetic field, which is assumed to
be constant within the envelope.
The Figure \ref{fig:tbts} gives the values of $T_s$ at $\Theta = 0^0
\; {\rm and} \; 90^0$ and we will use Equ.~\ref{equ:TsB} and \ref{equ:chi}
to define $T_s$ at intermediate angles, thus taking
\be
\chi_0 \equiv \frac{T_s(\Theta = \pi/2)}{T_s(\Theta = 0)}.
\label{equ:chi0}
\ee
Two main approximations are done in deriving Equ.~\ref{equ:TsB} and
\ref{equ:chi}.
The first one is that the radial temperature gradient $dT/dr$ is much
larger than the meridional temperature gradient $dT/dx$ ($x$ being a
coordinate along the star's surface): $T$ drops by about two orders of
magnitude across the envelope whose thickness is only a few tens of
meters, while it varies by less than a factor 10 (see Fig.~\ref{fig:tbts})
along the surface on length scales of kilometers, thus $dT/dr >>
dT/dx$.
The second approximation is that the thermal conductivities
$K_{\parallel}$ and $K_{\perp}$ are constant:
Hernquist \& Applegate (1984) showed that in fact, in the non magnetic
case, the temperature profile $T(\rho)$ within the outer part of
the envelope where the conductivity is dominated by the photons (the
region where most of the temperature drop occurs) does follow a
trajectory of constant $K$.
The validity of Equ.~\ref{equ:TsB} and \ref{equ:chi} can however only be
checked by comparison with two-dimensional calculations.
Schaaf (1990b) has performed such a calculation with relative success
and many approximations which make it far from definitive, but his
results are well reproduced by these simple formulas
Equ.~\ref{equ:TsB} and \ref{equ:chi}.
Awaiting a more detailed study, Equ.~\ref{equ:TsB} and
\ref{equ:chi} are the best that can be done.

\subsection{The atmosphere and the emerging spectrum} 

The structure of the atmosphere is of critical importance for
observations.
The total flux permeating it is controlled by the
underlying envelope but the spectral distribution of the flux
(i.e., the emerging spectrum) is determined in this region.
In the present work we will assume for simplicity that this emerging
spectrum is a blackbody spectrum.
Romani (1987) took into account various possible chemical compositions
of the atmosphere to calculate more realistic spectra in a zero
magnetic field approximation.
Due to the general $\omega^{-3}$ frequency dependence of the opacity
$\kappa(\omega)$ of a non magnetic fully ionized plasma, the specific
flux is increased at high frequencies, reduced at low frequencies, and
differs substantially from a blackbody spectrum.
When partial ionization is taken into account the excess in the Wien
part of the spectrum is reduced, the effect being most dramatic when
metals are present.
For most relevant temperatures, this part of the spectrum lies within
the {\em Einstein} and {\em ROSAT} detector ranges.
Inclusion of the magnetic effects is an enormous challenge and
extensive calculations are in progress (Shibanov \etal 1992;
Pavlov \etal 1994).
For a fully ionized magnetic plasma $\kappa^B(\omega) \sim
\omega^{-1}$ and the resulting spectrum is intermediate between the
blackbody and the fully ionized non magnetic spectrum.
Since the magnetic field increases substantially the electron binding
energies, hydrogen absorption edges are present even at high
temperature and alter the spectrum precisely within the
detector ranges, pushing it toward the Planck spectrum or
even below at low enough temperature (the details of these absorption
edges are unfortunately blurred by the PSPC's low energy resolution).

The blackbody approximation may finally not be as bad as it
appeared originally in Romani's work.
Moreover, the magnetic spectra depend on many parameters which have to
be added to the parameter set of the models presented here and will
make the analysis much more complicated.
For these reasons, in this first work, we will consider only blackbody
emission, characterized by the local effective temperature, and
reserve more realistic spectra for future work.

\section{GENERATING LIGHT CURVES AND SPECTRA \label{sec:litc_sp}}

\subsection{Flat Space-time \label{sec:flat}} 

Assume we have a neutron star at a distance $D$ from the observer and
we want to calculate the number of photons of energy $E$ reaching a
detector whose effective area for these photons is $A(E)$.
Choosing spherical coordinate $(\theta,\phi)$ on the star's surface
with the coordinate axis pointing toward the detector, the number of
photons emitted at the surface by an area
$d^2A = R^2 \: \sin \theta d\theta \: d\phi$,
where $R$ is the star's radius, is
\be
d^6 N =
     {\rm Phase \; Space \;\ Volume} \times {\rm Distribution \; Function}
       =
         2 \frac{d^3 x d^3p}{h^3}    \times       n(E,T;\ldots)
\ee
where the distribution function $n$ may depend, beside the photon
energy $E$ and the temperature $T$ of the emitting region $d^2A$, on
the local magnetic field, chemical composition, etc...  For blackbody
emission $n = 1/(e^{E/k_B \, T}-1)$.  Let's write
\be
d^3 x = c \, dt \cdot d^2A \cos \theta
\label{equ:d3x}
\ee
($c$ is the speed of light and $dt$ the time during which the received
photons have been emitted)
and
\be
d^3 p = p^2 \, dp \: d\Omega_p = \frac{E^2 dE}{c^3} \: \frac{A(E)}{D^2}
\label{equ:d3p}
\ee ($d\Omega_p = A(E)/D^2$ is the detector's solid angle as seen from the
star's surface and $p = E/c$ the photon momentum), and integrate the
emission over the portion of the star visible to the observer.
Interstellar absorption at energy $E$ is included through a factor
$\exp(-N_H \, \sigma(E))$, where $N_H$ is the effective hydrogen column
density between the star and the observer and $\sigma(E)$ the effective
cross section (Morrison \& McCammon 1983).
We obtain thus
\be
d^2 N(E) =
            \frac{2 \pi}{c^2 h^3} \, \frac{R^2}{D^2} <n(E;\ldots)>
            e^{-N_H \sigma(E)} E^2 \, A(E) \, dt \, dE
\ee
where
\be <n(E;\ldots)> \equiv \int_{0}^{1} 2x \, dx
                         \int_{0}^{2 \pi} \frac{d\phi}{2 \pi} \;
                         n(E,T(\theta,\phi);\ldots)
\label{equ:ave}
\ee
is the distribution function averaged over the visible part of the
star and $x \equiv \sin \theta$.

\subsection{Curved Space-time} 

General relativistic effects are twofold here: red-shift and lensing.
The red-shift makes that photons emitted at energy $E$ are received at
energy $E_{\infty} = E e^{\phi} < E$ while photons emitted during a
time $dt$ are received in a time $dt_{\infty} = e^{-\phi} dt > dt$
where
\be
e^{\phi} \equiv \sqrt{1-\frac{2 G M}{R c^2}} < 1.
\ee
With this we can rewrite the photon flux in term of observed quantities as
\begin{eqnarray}
d^2 N(E_{\infty}) = \frac{2 \pi}{c^2 h^3} \, \frac{R_{\infty}^2}{D^2}
                    <n(e^{-\phi} \, E_{\infty};\ldots)>
                    e^{-N_H \sigma(E_{\infty})} E_{\infty}^2 \,
                    A(E_{\infty})  \, dt_{\infty} \, dE_{\infty} \nonumber \\
              \equiv {\cal N}(E_{\infty})
                    A(E_{\infty})  \, dt_{\infty} \, dE_{\infty}
\label{equ:flux}
\end{eqnarray}
where
\be
R_{\infty} \equiv R e^{-\phi} = \frac{R}{\sqrt{1-\frac{2 G M}{R c^2}}} > R
\label{equ:rinf}
\ee
and the photon flux ${\cal N}(E_{\infty})$ has dimension
[${\rm photons \; cm^{-2} \; sec^{-1} \; keV^{-1}}$].

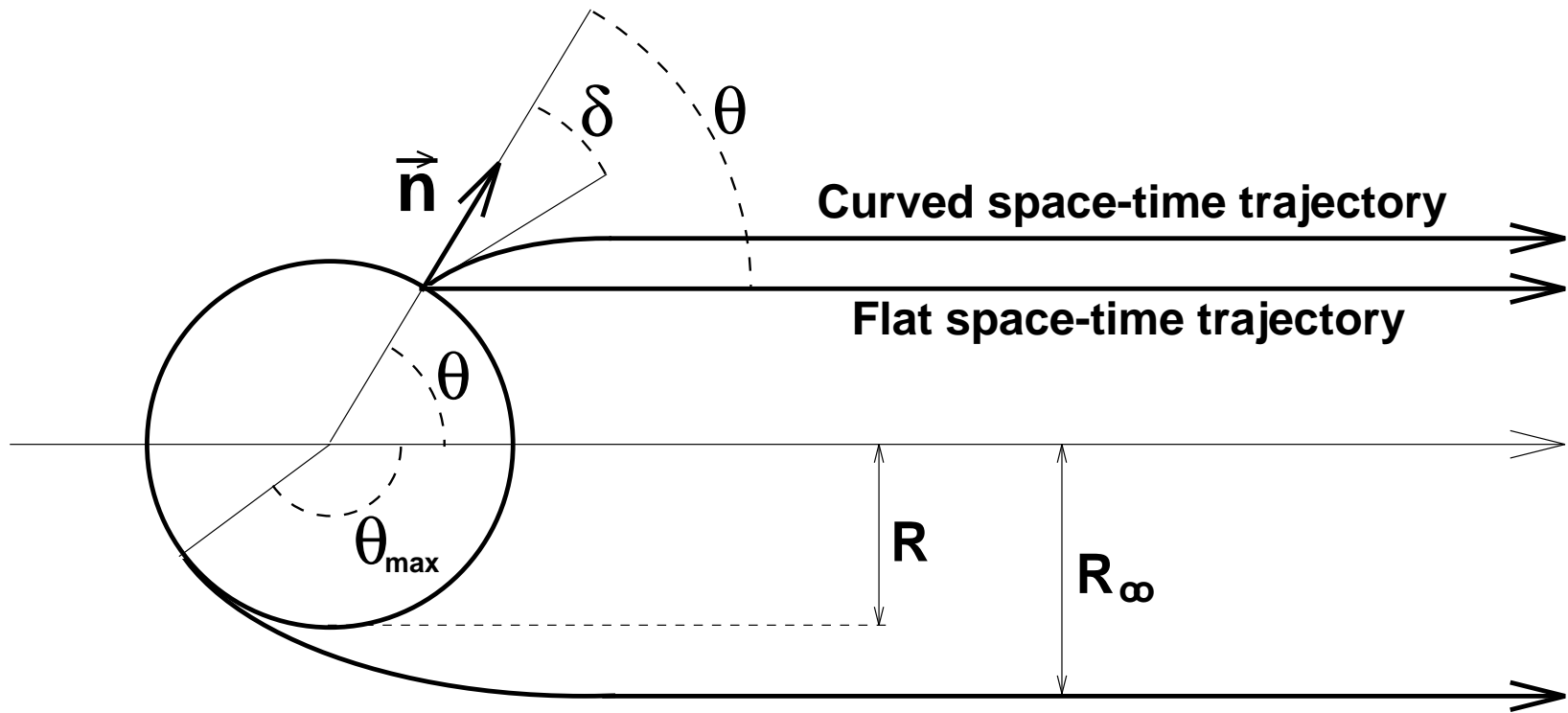
\begin{figure}
\caption{\label{fig:lens}
        Geometry of gravitational lensing.}
\end{figure}

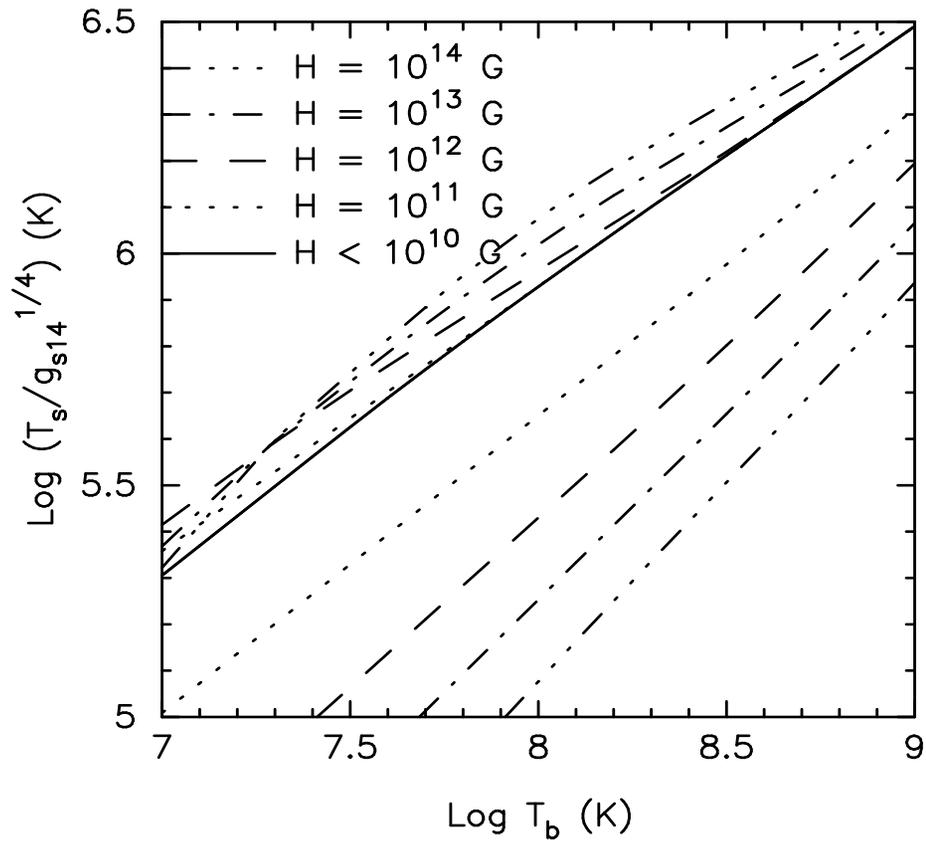
\begin{figure}
\caption{\label{fig:th_max}
        Maximum lensing angle vs. neutron star's radius for a 1.4 \Msol
        star and vs. radius in unit of the Schwarzschild radius
        $R_S = 2 G M / c^2 = 2.95 \; {\rm km} \; M/\Msol$.}
\end{figure}

The effect of gravitational lensing on photons emitted by hot spots on
the surface of neutron stars has been studied in detail by
Pechenick, Ftaclas \& Cohen (1983).
More than half of the star's surface is visible to an observer at
infinity (Fig.~\ref{fig:lens}) and a photon emitted at collatitude
$\theta$ on the star's surface which reaches the observer must be
emitted at an angle $\delta$ to the star's normal, defined through
\be
\theta = \theta (x) =
         \int_{0}^{GM/Rc^2} \frac{x \; du}
         {\sqrt{ \left(1-\frac{2 G M}{R c^2} \right)
          \left( \frac{GM}{Rc^2} \right)^2
           -(1 - 2u)u^2 x^2}}.
\label{equ:theta}
\ee
where
\be
x \equiv \sin \delta.
\label{equ:delta}
\ee
The impact parameter $b$ of the photon trajectory is then given by
$b = x \cdot R_{\infty}$.
In flat space-time $\delta = \theta$ while here $\delta < \theta$.
The maximum collatitude visible to an observer at infinity corresponds
to $\delta = \pi/2$, i.e., $\theta_{max} = \theta(x=1)$, and the star
appears as a disc of radius $R_{\infty}$.
For a given star mass, $R_{\infty}$ is minimal at $R = (3/2)R_S$,
with $R_S = 2GM/c^2 = 2.95 \; {\rm km} \; M/\Msol$ the star's
Schwarzschild radius, and is $R_{\infty}^{min} = (3\sqrt{3}/2)\, R_S
\simeq 2.6 R_S$
(thus, a 1.4 \Msol neutron star radius cannot appear smaller than
10.74 km, for an actual value of 4.134 km).
We show in Fig.~\ref{fig:th_max} the value of the maximum lensing
angle $\theta_{max}$ as a function of the star's radius $R$.
$\theta_{max}$ reaches $180^0$ at $R/R_S \simeq 1.76$, $360^0$ at
$R/R_S \simeq 1.5091$, and goes to infinity when $R/R_S$ tends to $1.5$
($r_{\gamma} = 1.5 R_S$ is the so-called `photon radius' at which a
photon can be in a circular Keplerian orbit arround the star, if $R <
r_{\gamma}$).

To obtain the flux received by a detector the only change compared to
the development of \S~\ref{sec:flat} is in Equ.~\ref{equ:d3x} where
$\cos \theta$ becomes $\cos \delta$ (the surface element $d^2 A$
remains $R^2 \: \sin \theta d \theta \: d\phi$ by definition of the
Schwarzschlid coordinates) and in the calculation of the detector's
solid angle $d\Omega_p$.
The latter can be seen to be (see, e.g., Appendix I in Pechenick
\etal 1983):
\be
d\Omega_p =
   \frac{x \; dx}{\cos \delta \: \sin \theta \: d \theta}
   \left[\frac{A(E)}{D^2}\right].
\label{equ:domega}
\ee
The term $x \; dx$ in Equ.~\ref{equ:ave} is
$\cos \theta \: \sin \theta \: d \theta$ which is exactly canceled by
the term $\cos \delta \: \sin \theta \: d \theta$ of Equ.~\ref{equ:domega}
(since $\cos \theta \rightarrow \cos \delta$ in Equ.~\ref{equ:d3x}).
The photon flux ${\cal N}(E_{\infty})$ is thus still given by
Equ.~\ref{equ:flux} and \ref{equ:ave} but now with $x \equiv \sin \delta$.

The actual sizes of neutron stars are unknown and this lensing effect
rises the hope of imposing some constraints on them.
Theoretical predictions, for a 1.4 \Msol star, range from 6.5 km up to
a maximum of 14 km.
The smallest value (6.5 km) appears in models with either a kaon
condensate (Thorsson, Prakash \& Lattimer 1994) or hyperons
(Pandharipande 1971) and the largest values (14 km) are obtained in
models with only nucleons and developed within the mean field theory.
For models  with only nucleons, the classical equation of state (EOS)
of  Friedman \&  Pandharipande  (1981)  and its  improved versions by
Wiringa, Fiks \& Fabrocini (1988) give radii between  10.5 and 11.5 km
at 1.4 \Msol.
Very low mass neutron stars have much larger radii but, when restricting
ourselves to masses arround 1.4 \Msol, 14 km should be considered as
an upper limit in the models.

\newpage   

\subsection{The detector's response \label{sec:det}}

The number of counts in the detector's output channel \#i, $Cts(i)$, is
obtained from ${\cal N}(E_{\infty})$ with the detector's response matrix
$(r_{ij})$:
\be
Cts(i) = \Delta t \; \sum_{j} r_{ij} \; {\cal N}(E_j) \; \delta E_j.
\label{equ:counts}
\ee
For {\em ROSAT}'s PSPC there are 256 output channels ($i = 1, \ldots
,256 = I$, but channels 1 - 7 have a zero effective area) and the
incoming flux is split in 729 energy bands $\delta E_j$ ($j = 1,
\ldots ,729 = J$).
$\Delta t$ is the observing time.
The $r_{ij}$ have dimension [$\rm cm^{-2}$] and the detector's effective
area at energy $E_j$ is $A(E_j) = \sum_i r_{ij}$ while the detector's
effective area for channel $i$ is $A_i = \sum_j r_{ij}$.
In all our calculations we use the 1992 March 19 release of the PSPC's
response matrix.
A perfect resolution detector would have $I=J$ and
$r_{ij} = \delta_{ij} \cdot A(E_i)$.

\subsection{Numerical method \label{sec:num}}

Given the star's mass $M$ and radius $R$ we define the magnetic
dipole's strength $B_p$ and inclination angle $\alpha$ with respect to
the rotation axis.
The magnetic field is calculated at each point of a fixed grid on the
star's surface (about one point per square degree) and, given the
interior temperature $T_b$, the surface temperature is calculated
on this grid using Equ.~\ref{equ:TsB} and \ref{equ:chi}.
We then define a mobile grid on the surface, which covers the part of the
star visible to the observer taking into account lensing, with usually one
point per $3^2$ square degrees or one point per square degree for cases of
very strong
lensing.
Given the observer's orientation $\zeta$ with respect to the rotation
axis, the mobile grid is rotated by steps of 6 degrees, at each step
the temperature is calculated on this grid by interpolating from the
values previously calculated on the fixed grid and the averaged
distribution function (Equ.~\ref{equ:ave}) is calculated, usually for
about one hundred values of the photon's energy $E$.
Given the observer's distance $D$ and the hydrogen column density
$N_H$, the fluxes ${\cal N}(E)$ are calculated (Equ.~\ref{equ:flux})
and then passed through the detector's response matrix
(Equ.~\ref{equ:counts}) to give the counts which can be directly
compared with the data.

\newpage  

\section{RESULTS \label{sec:results}}

\subsection{Surface Temperature Distributions \label{sec:Tsurf}}

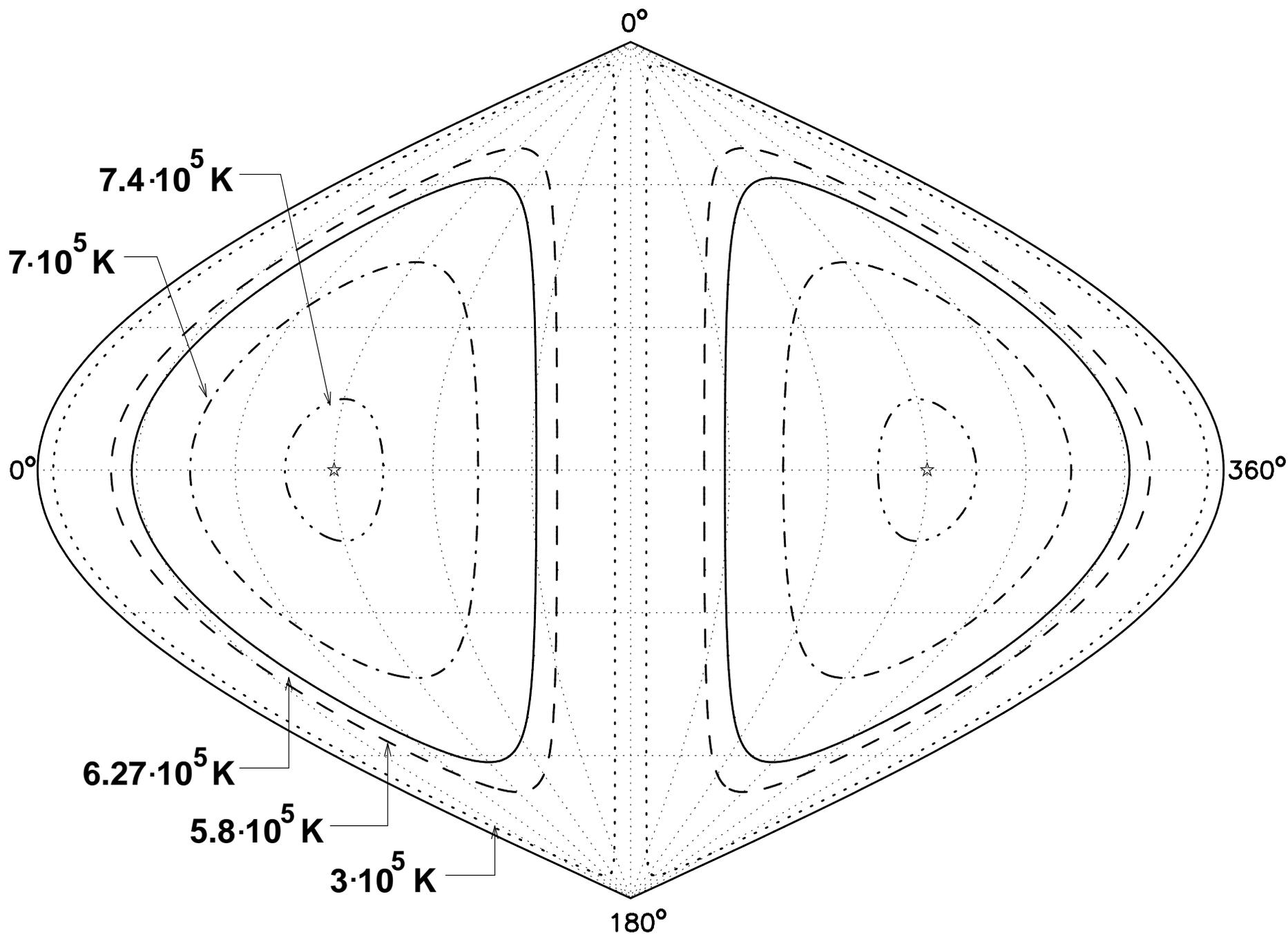
\begin{figure}
\caption{\label{fig:temp_dist}
         Surface temperature distribution with a dipolar magnetic
         field with field strength $B_p = 3 \cdot 10^{12}$ G at the poles
         (marked by stars) and an internal temperature of $6 \cdot 10^7$ K
         ($g_{s 14} = 1$).
         The effective temperature is $6.27 \cdot 10^5$ K and the average
         temperature is $5.8 \cdot 10^5$ K.
         The representation is an area preserving mapping.}

\end{figure}

With the method of \S~\ref{sec:tbts_H} we can now calculate surface
temperature distributions for given magnetic field configurations.
Figure \ref{fig:temp_dist} shows a typical example for a dipolar
surface field, other dipole strengths and internal temperatures giving
similar results.
Figures \ref{fig:t-a}a) and b) give the percentage of the star's area
with temperature inferior to a given value for various field
intensities ${B_p}$ and internal temperatures ${T_b}$.
For a given ${T_b}$, increasing ${B_p}$ results in a smaller
proportion of the surface having a temperature close to the minimal
temperature ${T_{min}}$, while for a given ${B_p}$ the same
effect is obtained by decreasing ${T_{b}}$.
Having the surface temperature distribution $T_s(\theta,\phi)$ we can
calculate the effective temperature $T_{e}=<T_s^4(\theta,\phi)>^{1/4}$
($<X>$ means surface average of $X$) and the average temperature
$T_{a}=<T_s(\theta,\phi)>$ which will be used below.
{}From Figures \ref{fig:temp_dist} and \ref{fig:t-a} one sees that most
of the surface is at a temperature close to the maximal temperature
$T_{max}$
(reached arround the magnetic poles), but that the temperature drops
to very low values on a small portion of the star.

\subsection{Spectra \label{sec:sp}}

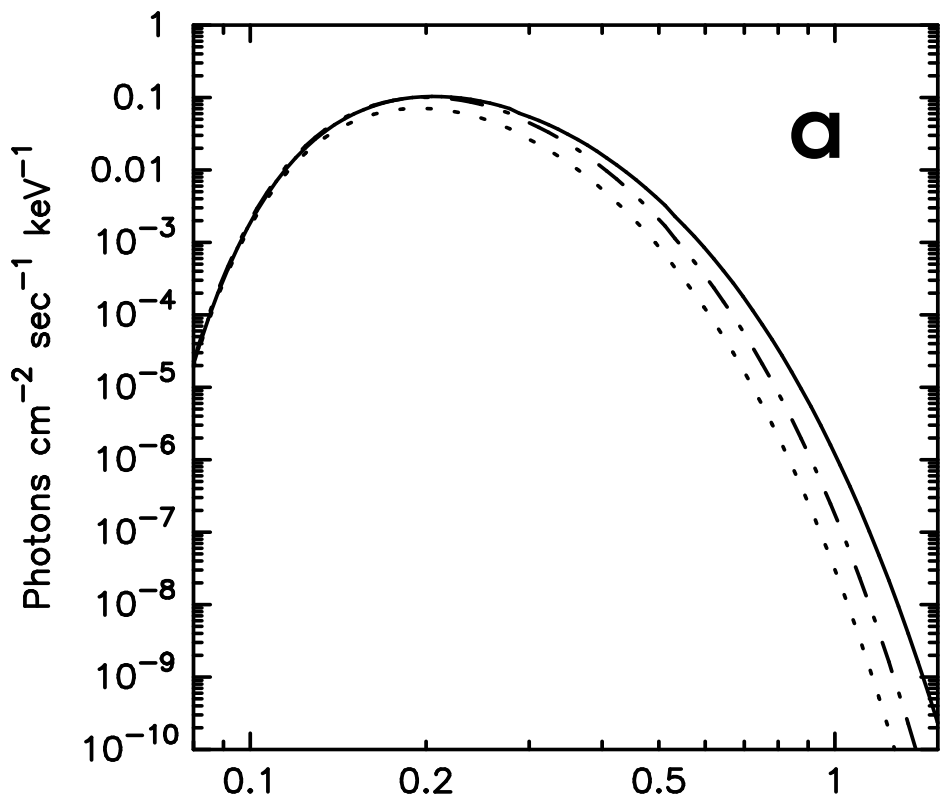
\begin{figure}
\caption{\label{fig:spec1}
        Incoming (a) and detected (b) spectra for the temperature
        distribution shown in Figure \protect\ref{fig:temp_dist}.
        ($D$ = 250 pc, $N_H = 10^{20} {\rm cm^{-2}}$ and
        $\alpha = \zeta = 90^0$).
        The continuous lines show the total composite spectrum
        while the dotted lines are a blackbody spectrum at the average
        temperature and the dashed lines a blackbody spectrum at
        the effective temperature.
        The incoming spectra (a) are ${\cal N}(E_{\infty})$ and the
        detected spectra (b) are $Cts(i)/\Delta t/\delta E$, i.e.,
        the incoming spectra passed through the response matrix.}
\end{figure}

In this paper we use blackbody emission at each point on the surface,
at the local temperature: our spectra are thus composite blackbody
spectra.
The effect of the non uniformity of the surface temperature on the
spectrum turns out to be relatively small, the resulting spectra being
very close to Planck spectra.
Figure \ref{fig:spec1} shows a typical example, the same dipolar field
as in Figure \ref{fig:temp_dist}:
one sees that the blackbody spectrum at the effective temperature
(dashed curves) is a reasonably good approximation to the `exact'
spectrum (continuous curves), particularly when the detector
resolution is taken into account where both spectra are almost
identical below 0.5 keV.
Since a second spectral component is present above 0.5 keV, spectral
fits to estimate the surface temperature are mostly sensitive to the
soft band below 0.5 keV and consequently, {\em as far as blackbody
fits are concerned, single temperature fits are practically as good as
our composite model to determine the effective temperature}.
Changing the field configuration and the observer's conditions
(orientation, distance and column density) makes little difference.
This result was already mentioned by Greenstein \& Hartke (1983).
The excess above 0.5 keV of the composite spectrum compared to the
blackbody spectrum at $T_e$ will though change the value of the
parameters used to model the hard component.

Phase dependent spectra are very close to the phase integrated
spectrum, the difference being often below the detection limit with
reasonable exposure times.
Again, changing the dipole's strength and orientation, or the
observer's conditions, makes little difference.
If spectra from magnetic atmosphere models were used the results may
be different since, when the temperature (and/or the surface field
strength) varies, the shape of the spectrum changes while the
blackbody spectrum's shape is invariant.
Gravitational lensing has obviously no effect on the phase integrated
spectrum, but will affect phase resolved spectra and make them even
closer to the phase integrated spectrum.
However light curves are more appropriate to study this effect.

\subsection{Light curves \label{sec:lc}}

The properties of the observed light curves (see Table~\ref{tab:data})
that one should keep in mind here are the presence of one or two
peaks, the phase of the peak(s) and the pulsed fraction, as well as
their dependence on the photon energy.
We define the pulsed fraction as:
\be
Pf(i) = \frac{\frac{1}{2}(Cts_{max} - Cts_{min})}{Cts_{mean}}
\ee
where $Cts_{max}$, $Cts_{min}$ and $Cts_{mean}$ are, respectively, the
maximal, minimal, and mean count rates in a given detector channel
(i=8,..., 256 for {\em ROSAT}'s PSPC)
The observed pulsed fractions are of course delicate to measure
and can only be defined for a range of channels, $Pf(i_1-i_2)$.

A surface dipolar field with two oppositely located warm regions on the
star's surface gives light curves with one or two peaks, and
singleness of the peak imply that the angle $\alpha$ between the
magnetic dipole and the rotation axis and the angle $\zeta$ between
the observer's direction and the rotation axis must satisfy, in flat
space-time,
\be
\alpha + \zeta < 90^0
\ee
to assure that only one magnetic pole is ever seen.
Gravitational lensing will impose a more stringent condition but, on
the other side, the statistical uncertainty in the observed counts
makes that low amplitude peaks in the light-curve cannot be
recognized.
We show in Figure~\ref{fig:litc0} seven examples of light curves with
$\alpha = \zeta$ and the corresponding pulsed fractions.
We find, for single peak configurations, that the maximum pulsed
fraction is obtained when $\alpha \sim \zeta \sim 45^0$, which is
obvious from geometrical reasons.
If $\alpha \sim \zeta > 45^0$ the pulsed fraction is comparable or
slightly lower than at $45^0$ and a second peak appears.
An important, simple, fact seen in {\em all} cases we have considered,
from which only the most representative are presented in this paper,
is that {\em the phase of the peak(s) in the light curve is
independent of the photon energy in the whole range 0.08 - 2.5 keV},
as can be seen for example in Fig.~\ref{fig:litc0}.

The lower the effective temperature $T_e$ the higher the pulsed
fraction $Pf$ which comes from the fact that at lower temperatures the
difference between $T_{max}$ and $T_{min}$ is larger (see
Fig.~\ref{fig:tbts}).
A surprising result is that the highest magnetic fields do not give
the largest pulsed fractions.
As shown in Fig.~\ref{fig:t-a}b,
a larger portion of the surface is at low temperature at field
strengths arround $3 \cdot 10^{11}$ G than at higher fields:
it is thus at $B_p \sim 3 \cdot 10^{11} - 10^{12}$ G that we obtain
the highest pulsed fractions, and stronger fields have a pulsed
fraction slightly lower.

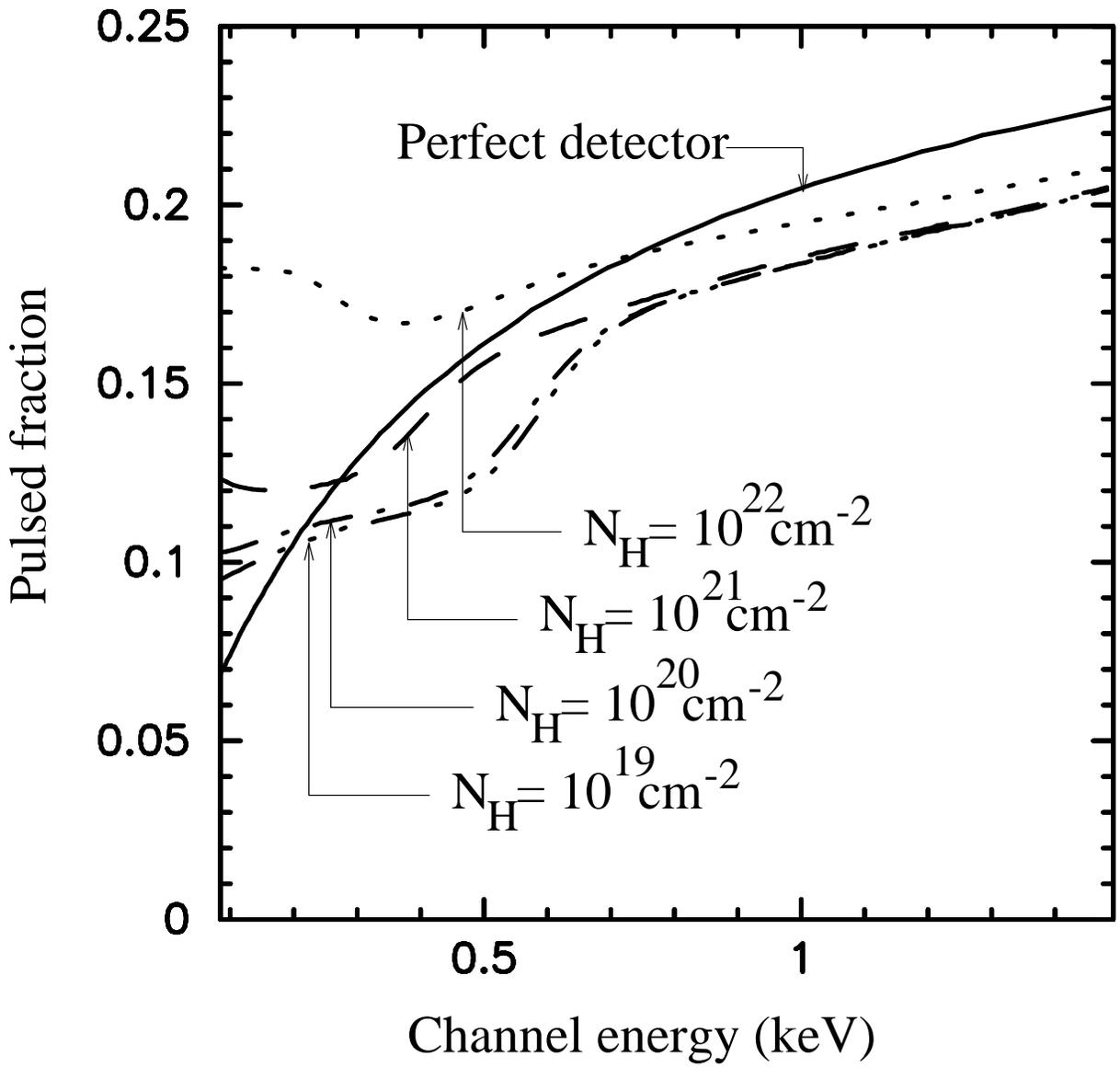
\begin{figure}
\caption{\label{fig:pls}
        Pulsed fraction for the configuration of
        Fig.~\protect\ref{fig:temp_dist} and \protect\ref{fig:spec1}
        but with different column densities $N_H$ as indicated.
        The continuous curve shows the pulsed fraction as would be
        obtained with a perfect detector, thus independent of the
        column density.}
\end{figure}

A general characteristic of pulsed fractions with dipolar fields and
blackbody emission is that they rise with photon energy, slowly at
low and high energy but with an intermediate region of faster increase
whose exact location depends on the effective temperature but is
always around 0.5 keV.
The presence of this steeper rise around 0.5 keV is actually a product
of the PSPC's response characteristics as can be seen from
Fig.~\ref{fig:pls} where the pulsed fraction of the incoming flux (as
it would be recorded by a detector with perfect energy resolution) is
compared with PSPC-detected pulsed fractions.
This feature will thus certainly be present even for more complex
surface temperature distributions.

If the detector has a very good energy resolution $Pf$ is
independent of interstellar absorption since its effect is just to
reduce the incoming flux at energy $E$ by a factor $\exp(-N_H \;
\sigma(E))$ which drops out when calculating $Pf$.
However, when dealing with a detector like {\em ROSAT}'s PSPC, the
finite energy resolution does affect the pulsed fraction by mixing in
the same output channel photons with different energies and thus
different absorptions.
Since $\sigma(E) \sim E^{-3}$, the net effect of interstellar
absorption is to suppress the soft photons compared to the harder ones
and when increasing $N_H$ the lowest PSPC's channels become
proportionally more and more contaminated by hard photons.
The result is a rise of the pulsed fraction in these lowest channels
(Figure~\ref{fig:pls}).
However, very large values of $N_H$ are required to obtain a
significant rise in $Pf$, and such high values are clearly excluded
in the cases of the four pulsars we consider:
they would dramatically reduce the flux at low energy, much below
anything observed.

The last point we consider is the effect of gravitational lensing.
The effect on the spectrum is very small (at least for dipolar fields)
and at the limit of observability due to the low count rates, but when
adding the output of many channels into light curves their flattening
can be clearly seen:
{\em the pulsed fraction is strongly dependent on the size of the
star}, besides its much weaker dependence on the dipole's strength and
the effective temperature.
We show in Figure~\ref{fig:lens_litc1} a sequence of light curves where the
star's radius is decreased from $\infty$ (i.e., no lensing effect) down to
6.25 km (i.e., $R/R_S = 1.51$), the star's mass being kept at 1.4 \Msol.
The corresponding pulsed fractions are shown in Fig.~\ref{fig:lens_pls1}.
The dipole and the observer are at $90^0$ to the rotation axis and the
dipole strength and interior temperature are choosen to maximalize the
pulsations according to the above discussion.
One immediately sees the dramatic effect of lensing which reduces $Pf$
from about 25\% (below 0.5 keV) in flat space-time to a half of this at
a radius of 18 km (an unrealistically large value for a 1.4 \Msol neutron
star) and to a tenth of this at a radius of 10 km.
As another example, we took a series of 6 neutron stars with masses and
radii of: (1.0 \Msol, 11.21 km), (1.2 \Msol, 11.08 km), (1.4 \Msol,
10.89 km), (1.6 \Msol, 10.55 km) and (1.8 \Msol, 9.76 km) (these radii
correspond to the EOS of Friedman \& Pandharipande 1981).
With an internal temperature of $5 \times 10^7$ K, a dipole strength of
$3 \times 10^{12}$ G, and $\zeta = \alpha = 90^0$, we obtained pulsed
fractions, in the energy band 0.08-0.5 keV, of 12\%, 8\%, 4\%, 1\%, and
1\% respectively.
There is also a tricky effect at radii between 8.85 km and 7.3 km: the
peak of the light curve is $90^0$ off-phase with respect to larger or
smaller radii.
At large radii ($R > 8.85$ km) the peaks correspond to one of the
magnetic poles being seen face-on and when $R$ decreases more of the
cool region in the magnetic equator is visible at this phase, thus
reducing the amplitude of the pulsations.
When $7.3 < R < 8.85$ km, the flux with the poles face-on is now lower
than when they are seen sideway (phases 0.25 and 0.75) where both warm
polar regions are wholly seen because of the strong lensing; at still
smaller radii, the second pole becomes visible in the face-on
configurations and the peak is back at phases 0 and 0.5.
This latter effect of increase of the pulsed fraction at small radii
when the whole surface of the star is visible and a hot region passes
on the antipode with respect ot the observer, `gravitational beaming',
has already been described by Pechenick \etal (1983) and the off-phasing
of the peak at slightly larger radii can also be seen in their figures.
Orienting the magnetic dipole and the observer at $45^0$ to the
rotation axis, $\alpha = \zeta = 45^0$, gives very similar results:
the light curves have the shapes shown in Fig.~\ref{fig:litc0}, the
pulsed fractions are observationally indinstinguishable from the ones
of Fig.~\ref{fig:lens_pls1} (for the same temperature distribution)
and the off-phasing of the peaks also happens at practically the same
star's sizes.
If we define a phase dependent effective temperature $T_e^{\Phi}$ from
the average of $T_s^4(\theta,\phi) \cdot \cos \delta$ over the part of
the star seen by the observer, i.e., an effective temperature that fits
the phase dependent spectra, then, for the cases of
Fig.~\ref{fig:lens_litc1}, $T_e^{\Phi}$ at the maximum exceeds
$T_e^{\Phi}$ at the minimum by about 13\% when there is no lensing, down
to 6\% at $R = 18$ km, 4.2\% at $R= 14$ km and 0.9\% at $R=10$ km.

\newpage  

\section{DISCUSSION \label{sec:disc}}

The magnetic envelope calculations we use to model the surface
temperature distribution tell us that the temperature of the warmer
regions, where the field is radial ($\Theta_B \sim 0$), is several
time higher than the temperature of the regions where the magnetic
field is tangential ($\Theta_B \sim 90^0$) as shown in
Fig.~\ref{fig:tbts}.
There are thus very large temperature differences on the surface but
the resulting effective temperature seen by an observer at infinity,
who averages over the part of the star he sees, varies little during
the star's rotation.
Variations are of the order of 5-10\% for observed pulsed fraction of
10-30\% below 0.5 keV and the observed spectrum varies thus little
with the rotational phase.
Magnetic effects on the emitted spectrum by themselves are important
and it is possible that when included in our model they may produce
more significant and interesting variations in the observed spectra.
These will hopefully be included in future work and for the time being
we will concentrate on the light curves.

The first noticeable characteristic of the observed light curves for
the four pulsars we study is their substantial pulsed fraction: the
results of Fig.~\ref{fig:lens_pls1} show that these values, between
$10 - 30\%$
below 0.5 keV, could be obtained with simple dipolar fields.
However the radii required, to avoid the flattening from gravitational
lensing, are larger than the present theoretical predictions if the
neutron star mass is assumed to be 1.4 \Msol.
Models with only nucleons in the core give the largest radii, and the
most sophisticated models (Wiringa \etal 1988; M\"{u}ther, Prakash
\& Ainsworth 1987) give radii between 10 - 12.5 km while other
predictions barely reach 14 km (Glendenning 1985 e.g.).
Models with meson condensates, hyperons or quarks give smaller radii
(Thorsson \etal 1994, Glendenning 1985, Glendenning 1992)
For any of these values, the pulsed fractions in the energy range 0.08
- 0.5 keV, for whatever values of the parameters ($B_p$, $T_b$,
$\alpha$ and $\zeta$), are below 10\%.
Analyses of the Vela pulsar glitches do favor large radii (Link,
Epstein \& Van Riper 1992; Alpar \etal 1993) but values above 15 km (for
a 1.4 \Msol neutron star) needed to give high enough $Pf$ are not
supported to date by any neutron star model.
Older calculations of neutron stars with solid neutron cores did give
radii of the order of 16 km (Pandharipande \& Smith 1975), but the
solidification of neutrons in neutron stars is not considered anymore
as realistic.
Taking neutron stars of lower masses raises the pulsed fraction, e.g.,
up to 12\% for a 1.0 \Msol star with radius of 11.2 km as seen in
\S~\ref{sec:lc} using a realistic EOS, but still below 10\% at masses above
1.2 \Msol.
So considering stars of mass lower than 1.4 \Msol is not sufficient to
obtain the
high observed pulsed fractions unless a very stiff EOS is used.
However, the values obtained are not much below the observed ones
unless the star's radius is small.

Spectra from magnetized hydrogen atmospheres give by themselves pulsed
fractions of the order of 10\% (Pavlov \etal 1994) for a dipolar field
with a uniform surface temperature and {\em no} lensing:
when used with the appropriate temperature distributions it is
reasonable to expect that they will increase $Pf$ compared to the
blackbody case, but the effect of the magnetic field considered in
this paper will probably dominate (we obtain a 25\% modulation without
lensing).
Using non magnetic hydrogen atmosphere spectra (Page, Zavlin, Pavlov
\& Shibanov, work in progress) one can get higher $Pf$, between 10-20\%
for standard masses and radii;
the increase, compared to blackbody spectra, being due to the slight
radial beaming produced by the limg darkening effect.

Our model shows that, for a fixed magnetic field configuration, the
difference between the maximum and minimum surface temperature and hence
the observed pulsed fraction increases with decreasing
temperature.
It may be interesting to notice (see Table~\ref{tab:data}) that this
trend is present within our very limited sample of four objects.
It may just be a coincidence or may indicate the occurence of a general
pattern of field evolution in which pulsars after a few thousands of years
develop similar surface field configurations.

In the cases of PSR 0833-45, 0656+14 and 1055-52 the observed pulsed
fractions are weakly energy dependent below 0.5 keV, which is almost
what we obtain.
However, the slight rise in $Pf$ in our results may be important since
this does not correspond exactly to the observations: the observed
pulsed fraction of PSR 1055-52 seems to be slightly decreasing with
increasing energy (below 0.5 keV).
In the case of Geminga, though, there is a significant decrease of $Pf$
in the band $0.28 - 0.5$ keV compared to its value below 0.28 keV:
{\em this feature is absolutely impossible to reproduce in our models}
(we doubt that more complex field configurations will alter this
conclusion).
This may be evidence for the inadequacy of blackbody emission and the
result of genuine atmospheric magnetic effects on the emitted spectrum.
In case of a very high column density (see Fig.~\ref{fig:pls}) we do
obtain a decrease of $Pf$ at low energy but it is much smaller than
what is observed in Geminga and the column densities required, $N_H >
10^{21} \rm cm^{-2}$, are at least an order of magnitude too high.

Even if we can somewhat reproduce the observed pulsed fractions we
do not score well on the shape of the light curves in the three
cases where they are single peaked (PSR 0656+14, 0630+178 and
1055-52).
A single-peaked light curve with a dipolar magnetic field restricts
the geometry to $\alpha + \zeta \lsim 90^0$ and the resulting
theoretical light curves have a long flat minimum and relatively
narrow peaks (Fig.~\ref{fig:litc0}):
this is opposite to the observations where the peaks are broad and the
light curves are above their mean more that 50\% of the time.
The use of magnetic spectra will change the shape of the light curves
but probably in the `wrong' direction: emission is beamed along the
field and one can expect that the peaks will be narrower.
We can safely conclude from this that the surface magnetic field of
these three stars is certainly not dipolar.

A simple way out of this problem is the addition of a small quadrupolar
component (Page 1994b) to broaden the peaks, keeping $\alpha + \zeta
\lsim 90^0$.
That may be a reasonable solution for PSR 0656+14 but for PSR~1055-52
and 0630+178 the situation is probably different.
Phenomenological description of the radio emission indicates (see
Table~\ref{tab:data}) that $\alpha \sim \zeta \sim 30^0$ or $8.2^0$ for
PSR~0656+14:
if we accept this geometry then only one polar region is seen and the
broadening of the soft X-ray light curve's peak should be achieved
with an alteration of the dipolar field.
However the small value of $\alpha$ and $\zeta$ estimated by Lyne \&
Manchester (1988), $8.2^0$, yields pulsed fractions of at most 1\%
(Fig.~\ref{fig:litc0}) with a dipolar field (no matter what the
strength of the dipole or the temperature is);
magnetic effects on the emitted spectrum will doubtly be sufficient to
raise this to the observed 14\%.
The $30^0-35^0$ inclination of Rankin (1993) or Malov (1990) give $Pf
\sim 8-10\%$ and could work once magnetic spectra are used.
A quadrupole component added to the dipole can also easily increase
$Pf$ though (Page 1994b).

In the case of PSR 1055-52 and 0630+178 interpretation of the the
radio and $\gamma$-ray emission, respectively, indicate that $\alpha
\sim \zeta \sim 90^0$.
A dipolar surface field will definitely give two peaks:
if the dipolar component is orthogonal to the rotation axis there
obviously must be {\em strong} higher order multipolar components, as
already mentioned by Halpern \& Ruderman (1993).
This case will be considered in detail in a forthcoming paper (Page
1994b).
The alternative interpretation of PSR 1055-52 as being an almost
aligned rotator (Malov 1990) was supported according to Malov (1989)
by the absence of pulsation in the X-ray band in the {\em Einstein}
observation (Cheng \& Helfand, 1983), an argument invalidated by the
{\em ROSAT} detection of pulsations.
With $\alpha \sim \zeta \sim 30^0$ (Malov 1990) we obtain a
pulsed fraction just below 10\% (Fig.~\ref{fig:litc0}) in the most
optimistic case ($B_p = 10^{12}$ G, low temperature and large radius
of 16 km).
Smaller radii will reduce $Pf$ well below the observed 11\% but, on
the other hand, magnetic effect on the emitted sectrum will certainly be
sufficient to give 11\%.
Thus the inclination angles proposed by Malov (1990) are probably
able to give pulsed fractions as observed, but the asymmetric shape of the
light curve again requires a non dipolar surface field.

The light curves of PSR 0833-45 (Vela) are more delicate to interpret.
The two {\em ROSAT} observations, separated by eight months, show two
somewhat different light curves where several narrow subpeaks have moved
in phase and amplitude.
By considering the two observations together one can distinguish two
peaks, the larger one about $210^0$ wide and the smaller one about
$100^0$ wide with an amplitude about one third the amplitude of the
large peak.
However the dip which separates the small peak may be due to
fluctuations and this small peak inexistent.
An orthogonal configuration, $\alpha \sim \zeta \sim 90^0$, as
suggested by analysis of the radio pulse, with a surface dipolar
field, gives two soft X-ray peaks but the large difference in
amplitude between the two observed peaks imply a distortion of the
field from the pure dipole.
An orientation with $\alpha \sim \zeta \sim 60^0 - 70^0$ (as prefered,
e.g., in outer gap models of its $\gamma$-ray emission, Romani \&
Yadigaroglu 1994) would give better results, i.e., peaks of different
amplitudes, but the peaks are definitely narrower than observed (and
again inclusion of magnetic atmosphere effects will make the peaks
still narrower).
If the small peak is rejected then we are in the same situation as
with PSR 0630+178 and 1055-52:
there must be strong quadrupolar, or higher order, components in the
magnetic field.

Our discussion assumed that the contamination of the soft X-ray
emission by the hard component is negligible:
this is true when blackbody emission is used for both components, and
also when magnetic atmosphere effects are included in the soft
component (Meyer \etal 1994).
Should the hard component contribute significantly to the soft band
emission our conclusion may be affected.

Several models of neutron stars with a pion or kaon condensate give
very small radii (Maxwell \& Weise 1976; Thorsson \etal 1994) and thus
lead to very small modulations in the light curves, in sharp
contradistinction to what is observed.
We thus have a new tool to test these models, but one should not consider
that our present results rule them out so far:
surface magnetic field configurations more complex than the simple
dipoles considered here, as well as magnetic effects on the emitted
spectrum, may lead to stronger pulsations and the question needs more
study (Page 1994b).
If neutron stars are actually as small as these models with meson
condensates claim, the off-phasing of the light curves with respect to
the viewing (Fig.~\ref{fig:lens_litc1}) at radii between 7.5 - 8.5 km
must induce us to use extreme care in relating the soft X-ray peak
phases with pulsations observed at other wavelengths.
This range of radii includes the calculated values for a 1.4 \Msol
neutron star in many models.

\section{CONCLUSIONS \label{sec:concl}}

The simple model presented here, with dipolar fields, shows several
features similar to the observed ones but with significant shortcomings.
We obtain pulsed fractions which are lower that the observed ones, unless
very large radii or very low masses are assumed.
The shape of the light curves do not match the observations,
indicating the presence of deviations from a purely dipolar field,
the strength of the required correction depending on the assumed geometry
and the size of the star.
One can, however, reasonably expect that more complicated field
structure will be able to reproduce the observed pulsed fractions and
the shapes of the light curves: there is probably no need to invoke
external heating of the surface and/or magnetospheric absorption (see
Introduction) to reproduce the data, but this does not mean that these
processes are not present.
`Energy dependent pulse shapes and phases are trying to tell us the whole
story; we have to interpret them' (\"{O}gelman 1993), but deciphering the
story will not be easy.
A complete model will have to consider complex surface fields and
also include the magnetic effects in the atmosphere.

\acknowledgments 

I am gratefull to J.~H. Applegate, J.~P. Halpern,
Y. Shibanov and G.~G. Pavlov for discussions,
as well as A. Sarmiento, A. Serrano and E. Vazquez.
My interest in this problem was triggered by an unpublished talk of
S. Tsuruta (U. of Oklahoma, March 1991).
This work was supported by a HEA-NASA grant NAGW 3075 at Columbia
and by a DGAPA grant IN104092 at the UNAM.
This is contribution number 350 of the IA-UNAM.

\clearpage 

\hoffset=-0.6in
\begin{small}

\begin{planotable}{ccccc}
\tablewidth{0pt}
\tablecaption{\label{tab:data}
              Some observational properties \tnm{a}.}
\tablehead{               &       PSR 0833-45       &       PSR 0656+14       &
     PSR 0630+178       &       PSR 1055-52       \\
                          &         (Vela)          &                         &
       (Geminga)        &                         }
\startdata 

 $P$       (msec)         &           89            &           385           &
          237           &           197           \nl
 $\tau$    (yrs)          &   $1.2 \times 10^4$     &   $1 \times 10^5 $      &
    $3 \times 10^5$     &      $5 \times 10^5$    \nl
 \nl
 $T_e^{BB}$ (K) \tnm{b}   &  $1.5-1.6 \times 10^6$  &  $8.6-9.4 \times 10^5$  &
  $4.2-6.2 \times 10^5$ &  $6.3-7.5 \times 10^5$  \nl
$N_H^{BB}$(${\rm cm^{-2}}$)  \tnm{b}
                          & $0.1-1.5 \times 10^{20}$&$0.8-1.2 \times 10^{20}$ &
$ 1-2  \times 10^{20}$  & $2.0-4.5 \times 10^{20}$\nl
 Soft component \tnm{b}   &
                            $\begin{array}{c}
                            Pf \sim 11\% \nal
                   {\rm one \; or \; two \; peaks \; (?)}
                               \end{array}$
                                                    &
                                                        $\begin{array}{c}
                                                        Pf \sim 14\%  \nal
                                                        {\rm one \; peak}
                                                           \end{array}$
                                                                              &

    $\begin{array}{c}

      Pf \sim \left\{

     \begin{array}{c}

  33\% \; (0.08-0.28 keV) \nal

    10\% \; (0.28-0.5 keV)

       \end{array}

       \right. \nal

    {\rm one \; peak}

     \end{array}$

                        &

                           $\begin{array}{c}

                           Pf \sim 11\% \nal

                           {\rm one \; peak}

                             \end{array}$           \nl
 Hard component \tnm{b}   &        Unpulsed         &            ?            &
  $Pf \sim 35\%$        &$Pf\sim 20\rightarrow 80\%$ \nl
 $\phi_S - \phi_H$ (deg) \tnm{b}
                          &           n/a           &           n/a           &
     $\sim 105$         &    $\sim 120$           \nl
 \nl
 Radio pulsar             &    Yes (one pulse)      &    Yes (one pulse)      &
         No             &  Yes (two pulses)       \nl
 Optical counterpart      &    Yes   (pulsed)       &  Yes (proposed) \tnm{c}
& Yes (confirmed) \tnm{d} &   None to date         \nl
 $\gamma$-ray pulsar      &    Yes (two pulses)     &       No                &
  Yes (two pulses)      & Yes (one pulse) \tnm{e} \nl
 \nl
 $B_p$ (G)                &  $6.8 \times 10^{12}$   &  $8.8 \times 10^{12}$   &
 $3.2 \times 10^{12}$   &  $2.2 \times 10^{12}$   \nl
 $\alpha$ (deg) \tnm{f}   & \bba \LM: 90 \nal \R: 90 \nal \M: 35 \eab
                                           & \bba \LM: 8.2 \nal \R 30 \nal \M
35 \eab
                                                                              &
        $\sim 90$

        & \bba \LM: 74.7 \nal \R 90 \nal \M 30 \eab \nl
 $\zeta$ (deg)  \tnm{f}  & \bba \LM: 83 \nal \R: 78 \eab
                                           & \bba \LM: 8.2 \nal \R: 30 \nal \M:
35 \eab
                                                                              &
        $\sim 90$

       & \bba \LM: 66.8 \nal \R: 90 \nal \M: 30 \eab \nl
\tnt{a}{The rows list respectively: the period $P$,
                                    spin down age $\tau = P/2 \dot{P}$,
                                    effective temperature from blackbody fit
$T_e^{BB}$ and
                                    the associated hydrogen column density
$N_H^{BB}$,
                                    pulsed fraction $Pf$ and number of peaks of
the soft X-ray component,
                                    pulsed fraction $Pf$ of the hard X-ray
component,
                                    phase difference between the peaks of the
soft and hard X-ray components,
                                    detections at radio, optical and
$\gamma$-ray wavelengths,
                                    the dipolar field strength
(Equ.\protect\ref{equ:B_p}) and
                                    the estimated angles between the magnetic
dipole and the rotation axis ($\alpha$)
                                    and between the observer's direction and
the rotation axis ($\zeta$).}
\tnt{b}{Values are from \protect\"{O}gelman \protect\etal 1993 for Vela,
                        Finley \protect\etal 1992 for 0656+14,
                        Halpern \protect\etal 1993 for Geminga and
                        \protect\"{O}gelman \protect\etal 1993 for 1055-52.}
\tnt{c}{Caraveo \protect\etal 1994}
\tnt{d}{Bignami \protect\etal 1993}
\tnt{e}{Fierro \protect\etal 1993}
\tnt{f}{Values are from Lyne \protect\etal 1988 (LM),
                        Rankin 1993 (R),
                        Malov 1990 (M) and for
                        Geminga: Halpern \protect\etal 1993.}

\end{planotable}

\end{small}

\clearpage 


\clearpage

\begin{figure}
\caption{\label{fig:tbts}
        $T_b - T_s$ relationship for transport parallel and perpendicular to
        the field at various field strengths.
        $T_b$ is the temperature at the bottom of the envelope whose density
        is fixed at $10^{10} {\rm gm \; cm^{-3}}$ and $T_s$ is the `surface'
        or effective temperature.
        The dependence on the star's mass and radius is entirely contained
        in the scaling factor $g_{s \; 14}^{1/4}$.
        The continuous line gives the non magnetic relationship.
        The high values of $T_s$ correspond to a radial magnetic field
        (parallel transport) and are taken from Hernquist (1985) while
        the low $T_s$ values correspond to a field tangential to the
        surface (orthogonal transport) and are from Schaaf (1990a).
        The values of $T_s$ lower than $10^{5.5}$ K are linear extrapolations
        of the higher values and must be considered only as illustrative
        (see text).}
\end{figure}

\bigskip

\bigskip

\bigskip

\bigskip

\begin{figure}
\caption{\label{fig:t-a}
         Percentage of the star's area with temperature lower than a given
         value for dipolar fields.
         $T_{min}$ and $T_{max}$ are the minimum and
         maximum temperatures on the surface which can be read from
         Figure \protect\ref{fig:tbts}.
         \newline a) $B_p = 3 \cdot 10^{12}$ G with four different
         interior temperatures: $3 \cdot 10^7$ K (dash-dot),
                                $6 \cdot 10^7$ K (dash-triple dot),
                                $10^8$ K (continuous) and
                                $3 \cdot 10^8$ K (dash).
         \newline b) $T_{int} = T_b = 10^8$ K with five different
         dipole strengths: $3 \cdot 10^{11}$ G (dash-dot),
                           $10^{12}$ G (dash-triple dot),
                           $3 \cdot 10^{12}$ G (continuous),
                           $ 6 \cdot 10^{12}$ G (dash) and
                           $10^{13}$ G (dot).
         \newline
         These curves are independent of the mass and radius of the star.}
\end{figure}

\bigskip

\bigskip

\begin{figure}
\caption{\label{fig:litc0}
        Light curves (left) and pulsed fractions (right) for seven
        orientations of the observer and magnetic dipole:
        $\alpha = \zeta = 8^0, 15^0, 30^0, 45^0, 60^0, 75^0$ and $90^0$.
        In all cases the
        field is dipolar with strength $B_p = 10^{12}$ G and the
        internal temperature $T_b = 5 \times 10^7$ K;
        $T_e = 5.2 \times 10^5$~K.
        1.4 \Msol neutron star with a radius of 16 km,
        $N_H = 10^{20} \rm cm^{-2}$.
        The observer's distances are adjusted to avoid superposition
        of the light curves
        (the shape of the light curves and the value of the pulsed
        fraction are independent of the distance).}
\end{figure}

\bigskip

\bigskip

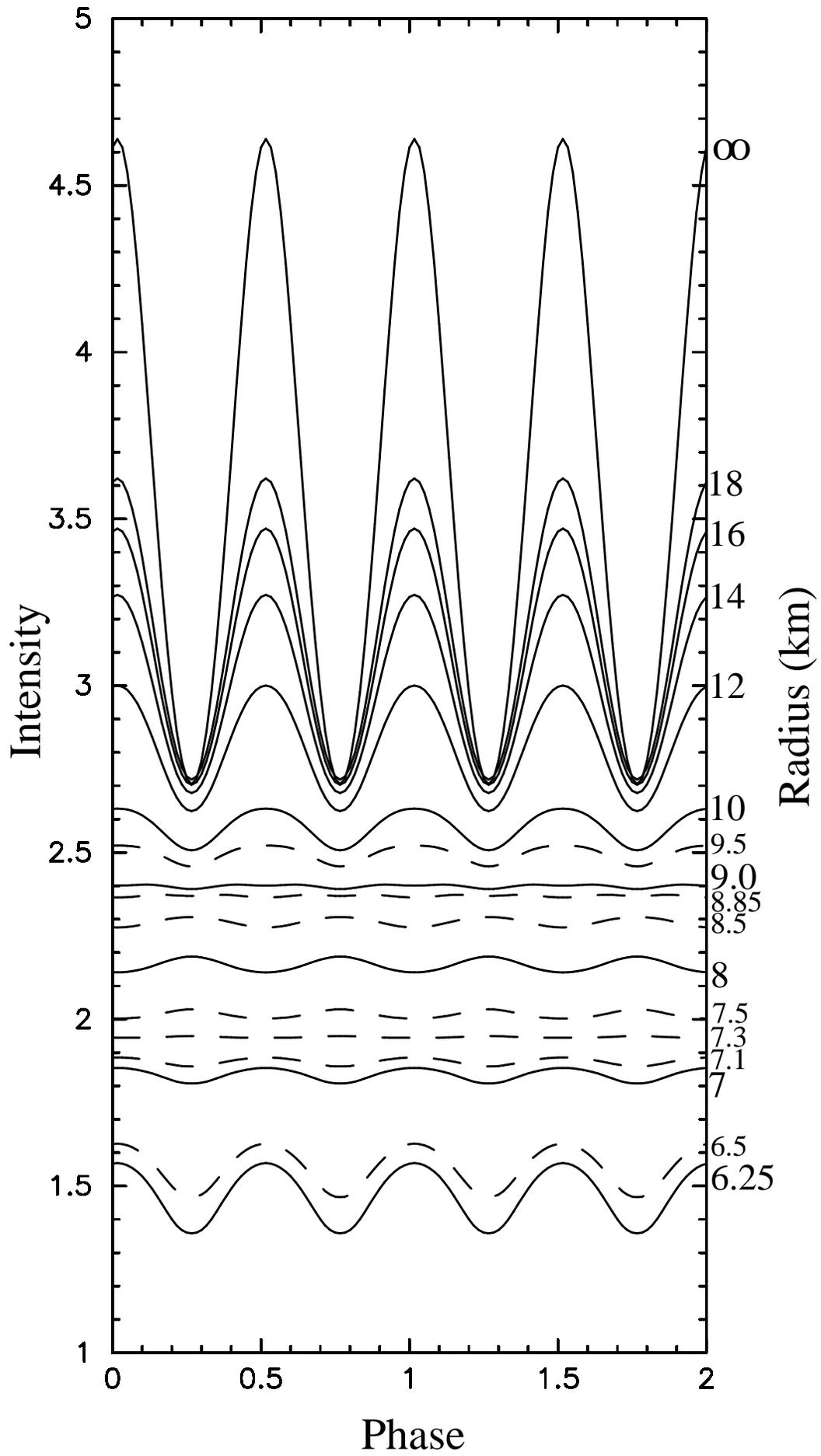
\begin{figure}
\caption{\label{fig:lens_litc1}
        Gravitational lensing of the light curves: orthogonal rotator.
        Sequence of light curves from a 1.4 \Msol neutron star with
        decreasing radii from $\infty$ (i.e., no general relativistic
        effects) down to 6.25 km.
        Since the surface temperature, for a given field strength and
        internal temperature, is proportional to $g_{s \: 14}^{1/4}$
        (see Fig.~\protect\ref{fig:tbts}) we have forced $g_{s \: 14}
        = 1$ in calculating the surface temperature distribution: all
        stars have exactly the same $T_s(\theta,\phi)$ and the effects
        shown are exclusively due to lensing and red-shift.
        We took $T_b = 5 \times 10^7$ K, $B_p = 10^{12}$ G and
        $\alpha = \zeta = 90^0$.
        The distances of the stars are chosen such that the total
        received flux is the same for all stars
        and $N_H = 10^{20} \; {\rm cm^{-2}}$.
        Lowering of the intensity for decreasing radii is due to the
        red-shift which slowly drives the flux out of the PSPC's range.
        Notice that for radii between 8.85 km and 7.3 km the peaks
        are $90^0$ off-phase with respect to the other cases.
        Phase 0 corresponds to the dipole pointing toward the observer.}
\end{figure}

\bigskip

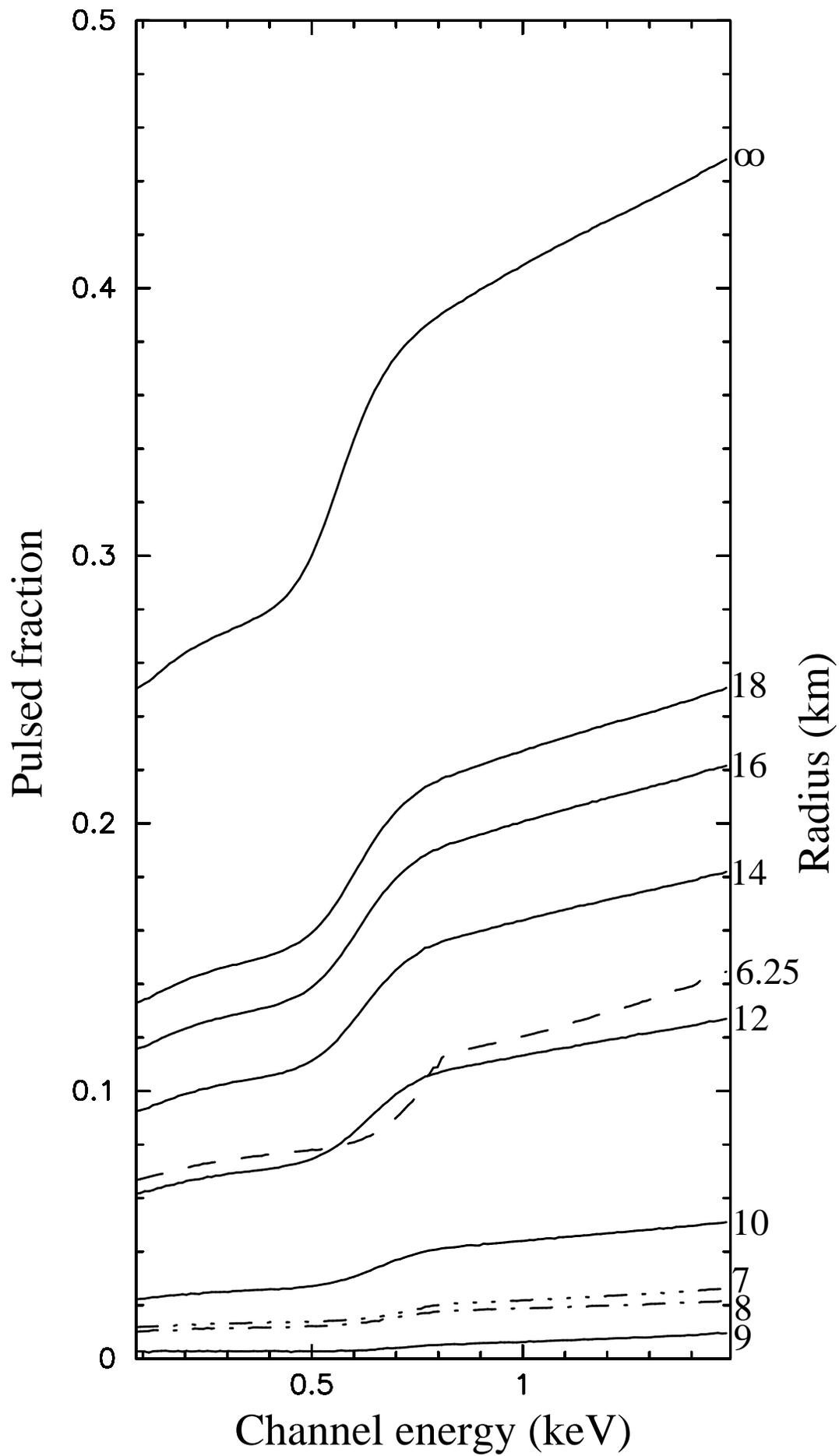
\begin{figure}
\caption{\label{fig:lens_pls1}
        Gravitational lensing of the pulsed fraction: same model as
        Fig.~\protect\ref{fig:lens_litc1} with radii
        from $\infty$ down to 6.25 km.}
\end{figure}

\bigskip

\clearpage

\begin{flushleft}
\bf
Author's Address
\end{flushleft}
 \noindent
Dany Page: Instituto de Astronom\mbox{\'{\i}}a, U.N.A.M.,\\
           Apdo postal 70-264 \\
           04510 MEXICO D.F., M\'{e}xico \\
\noindent
E-mail : PAGE@ASTROSCU.UNAM.MX\\

\end {document}
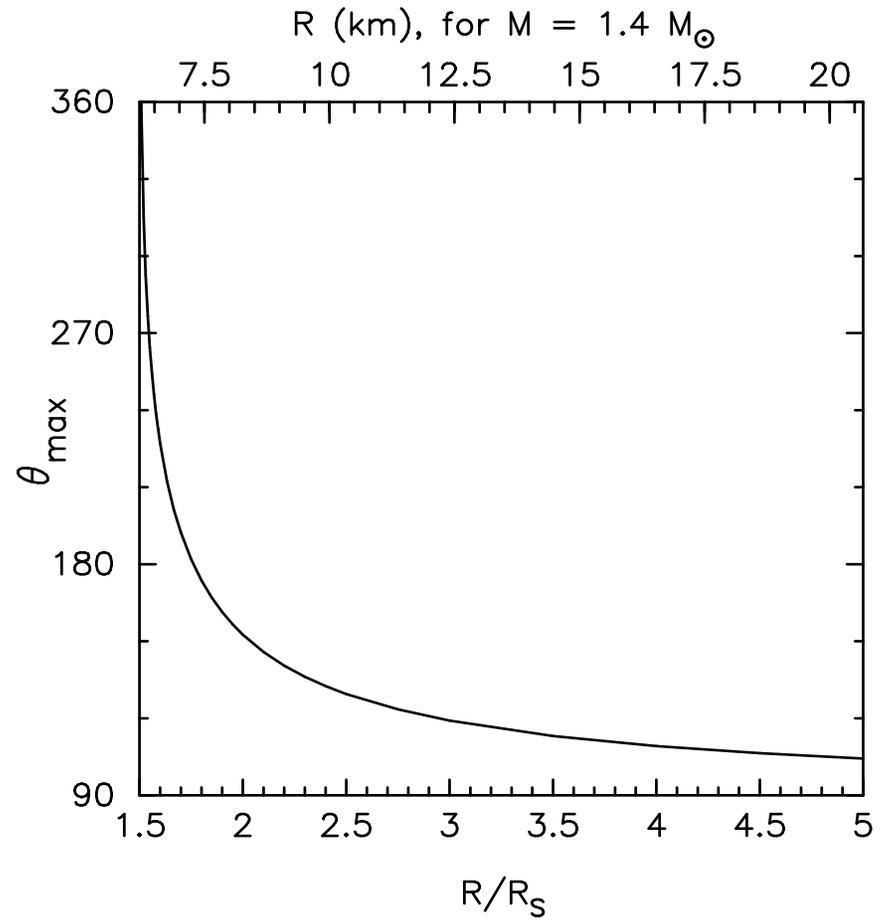
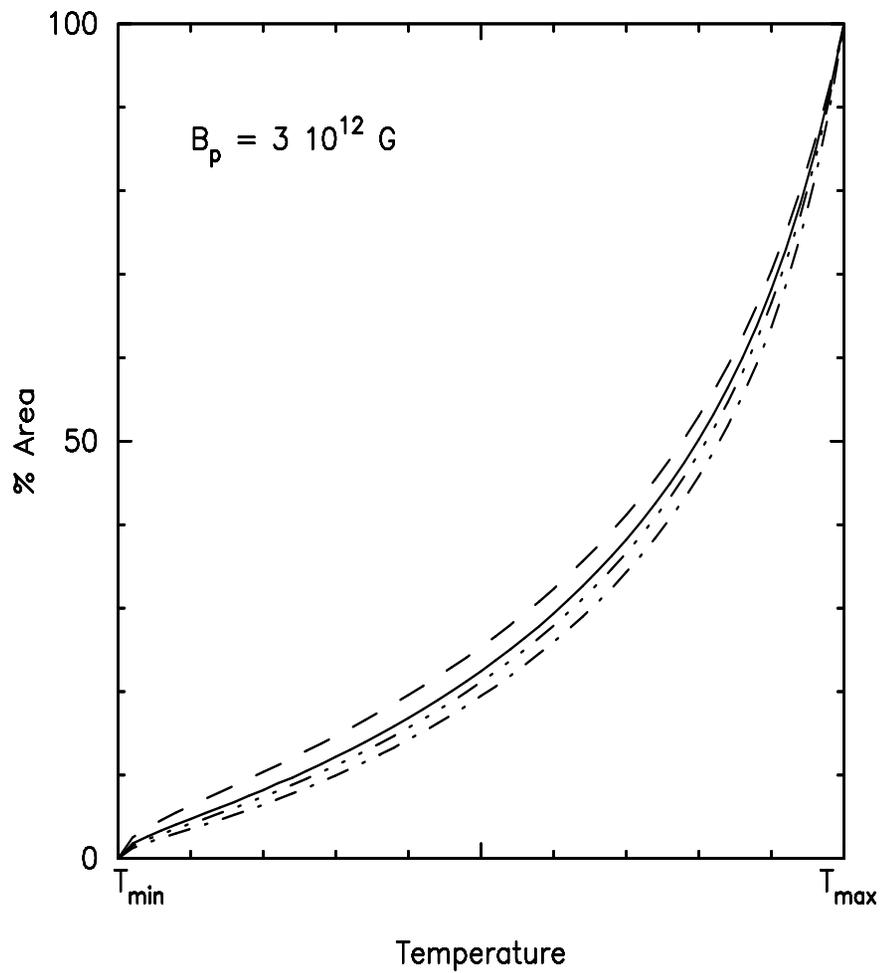
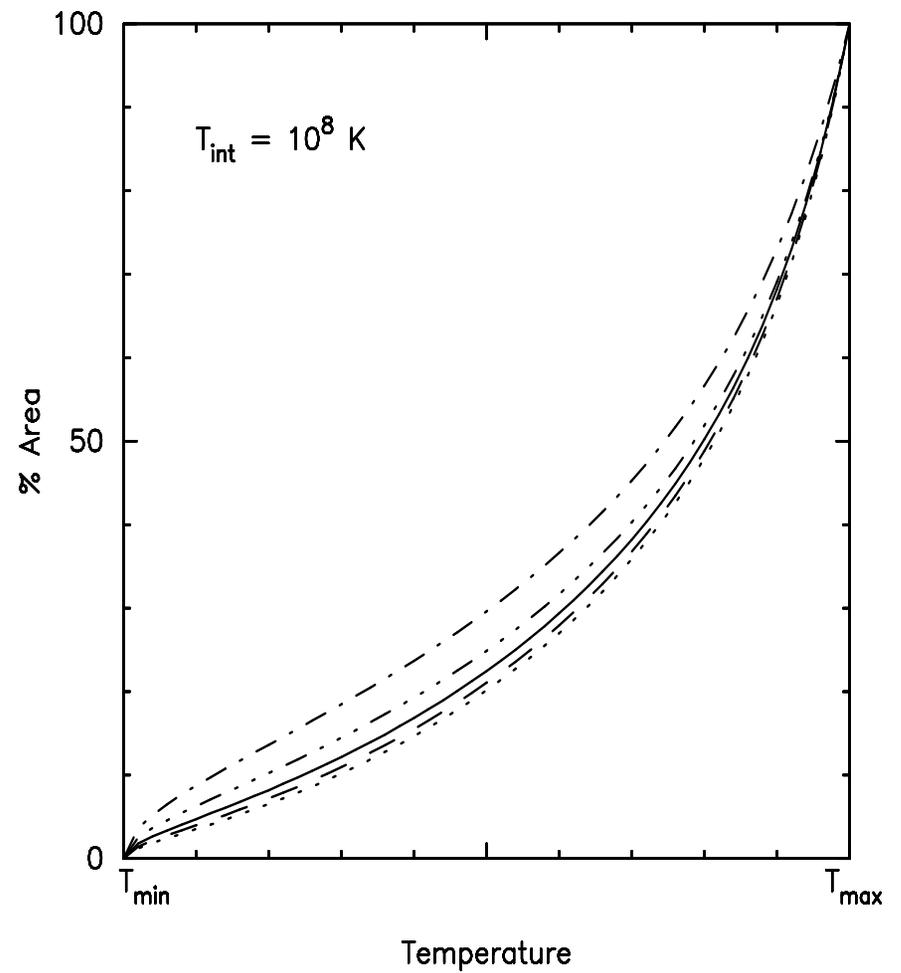
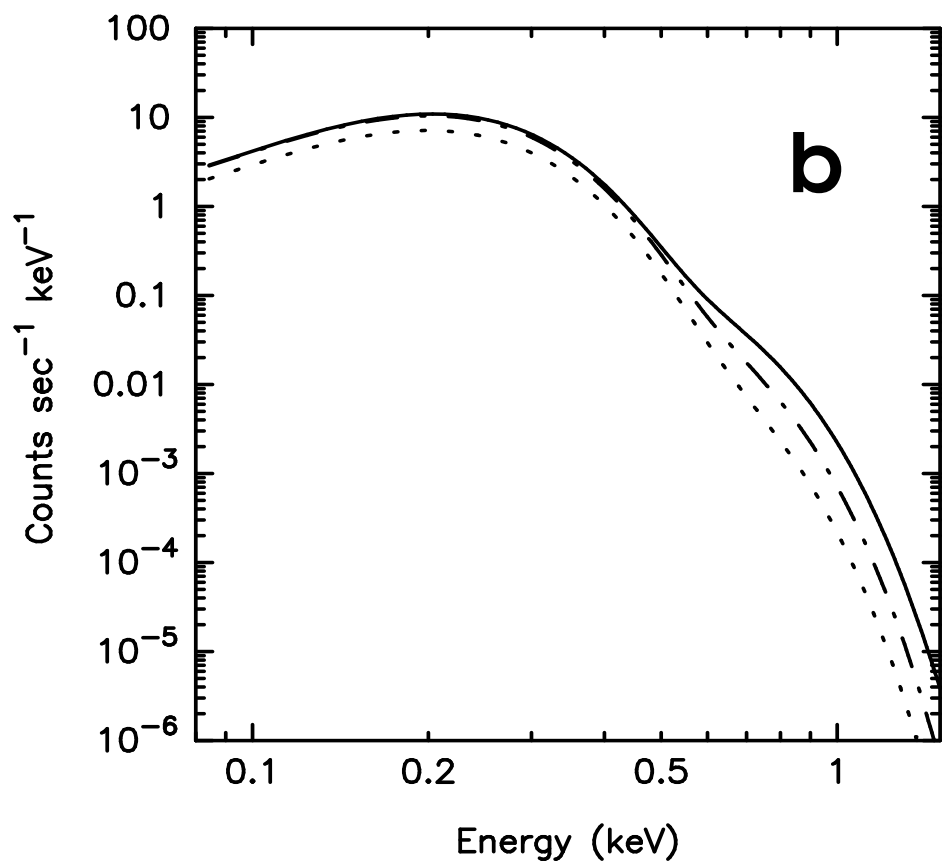
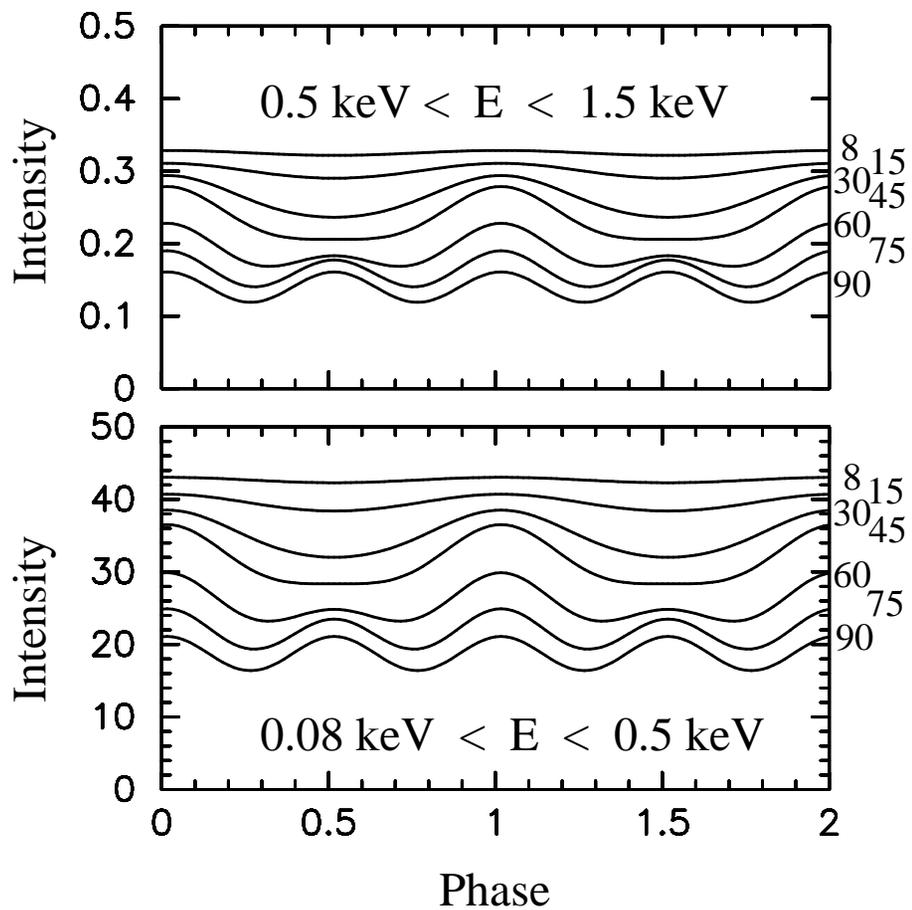
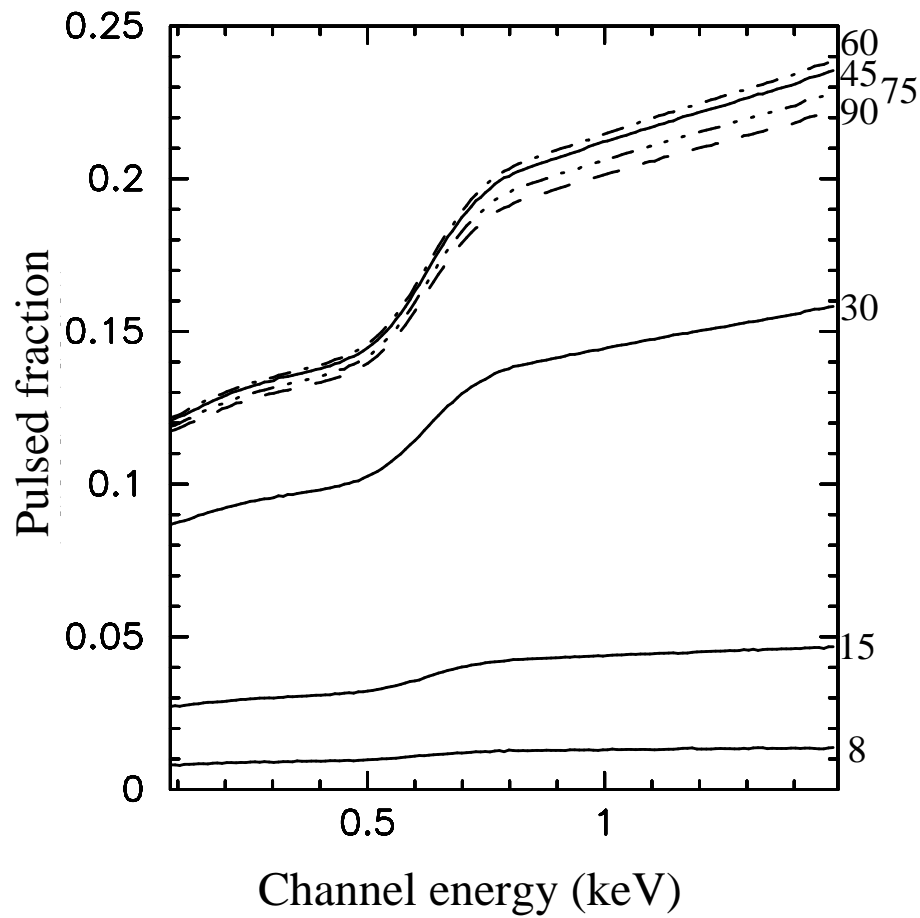
